\documentclass[12pt,usenatbib]{article}

\usepackage{amsthm,amsmath,natbib}
\usepackage{soul, color, multirow, graphicx}
\RequirePackage[colorlinks,citecolor=blue,urlcolor=blue]{hyperref}
\usepackage[text={7.1in,8in}, top=1.75in, left=0.69in]{geometry}

\newcommand{\cp}[1]{\texttt{#1}}
\newcommand{\beginsupplement}{%
        \setcounter{table}{0}
        \renewcommand{\thetable}{S\arabic{table}}%
        \setcounter{figure}{0}
        \renewcommand{\thefigure}{S\arabic{figure}}%
     }

\begin{document}

\title{Dynamic Prediction for Multiple Repeated Measures and Event Time Data: An Application to Parkinson's Disease}
\vspace{10mm}

\author{Jue Wang, Sheng Luo\footnote{Corresponding author: Sheng Luo is Associate Professor, Department of Biostatistics, The University of Texas Health Science Center at Houston, 1200 Pressler St, Houston, TX 77030, USA (E-mail: sheng.t.luo@uth.tmc.edu; Phone: 713-500-9554).}, Liang Li}

\date{}
\maketitle

\begin{abstract}
In many clinical trials studying neurodegenerative diseases such as Parkinson's disease (PD), multiple longitudinal outcomes are collected to fully explore the multidimensional impairment caused by this disease. If the outcomes deteriorate rapidly, patients may reach a level of functional disability sufficient to initiate levodopa therapy for ameliorating disease symptoms. An accurate prediction of the time to functional disability is helpful for clinicians to monitor patients' disease progression and make informative medical decisions. In this article, we first propose a joint model that consists of a semiparametric multilevel latent trait model (MLLTM) for the multiple longitudinal outcomes, and a survival model for event time. The two submodels are linked together by an underlying latent variable. We develop a Bayesian approach for parameter estimation and a dynamic prediction framework for predicting target patients' future outcome trajectories and risk of a survival event, based on their multivariate longitudinal measurements. Our proposed model is evaluated by simulation studies and is applied to the DATATOP study, a motivating clinical trial assessing the effect of deprenyl among patients with early PD.
\end{abstract}


\newpage
\section{Introduction}\label{sec:intro}
Joint models of longitudinal outcomes and survival data have been an increasingly productive research area in the last two decades \citep[e.g.,][]{Tsiatis:2004StatSinica}. The common formulation of joint models consists of a mixed effects submodel for the longitudinal outcomes and a semiparametric Cox submodel \citep{Wulfsohn1997Biometrics} or accelerated failure time (AFT) submodel for the event time \citep{Tseng2005Biometrika}. Subject-specific shared random effects \citep{Vonesh2006SIM} or latent classes \citep{Proust2014SMMR} are adopted to link these two submodels. Many extensions have been proposed, e.g., relaxing the normality assumption of random effects \citep{Brown2003Biometrics}, replacing random effects by a general latent stochastic Gaussian process \citep{Xu2001JRSSC}, incorporating multivariate longitudinal variables \citep{Chi2006Biometrics}, and extending single survival event to competing risks \citep{Elashoff2007SIM} or recurrent events \citep{Sun2005JASA,Liu2009JRSSC}.

Joint models are commonly used to provide an efficient framework to model correlated longitudinal and survival data and to understand their correlation. A novel use of joint models, which gains increasing interest in recent years, is to obtain dynamic personalized prediction of future longitudinal outcome trajectories and risks of survival events at any time, given the subject-specific outcome profiles up to the time of prediction. For example, \cite{Rizopoulos2011Biometrics} proposed a Monte Carlo approach to estimate risk of a target event and illustrated how it can be dynamically updated. \cite{Taylor2013Biometrics} developed a Bayesian approach using a Markov chain Monte Carlo (MCMC) algorithm to dynamically predict both the continuous longitudinal outcome and survival event probability. \cite{Blanche2015Biometrics} extended the survival submodel to account for competing events. \cite{Rizopoulos2013arXiv} compared dynamic prediction using joint models v.s. landmark analysis \citep{VanHouwelingen2007SJS}, an alternative approach for dynamically updating survival probabilities. A key feature of these dynamic prediction frameworks is that the predictive measures can be dynamically updated as additional longitudinal measurements become available for the target subjects, providing instantaneous risk assessment.

Most dynamic predictions via joint models developed in the literature have been restricted to one or two longitudinal outcomes. However, impairment caused by the neurodegenerative diseases such as Parkinson's disease (PD) affects multiple domains (e.g., motor, cognitive, and behavioral). The heterogeneous nature of the disease makes it impossible to use a single outcome to reliably reflect disease severity and progression. Consequently, many clinical trials of PD collect multiple longitudinal outcomes of mixed types (categorical and continuous). To properly analyze these longitudinal data, one has to account for three sources of correlation, i.e., inter-source (different measures at the same visit), longitudinal (same measure at different visits), and cross correlation (different measures at different visits) \citep{OBrien2004JRSSC}. Hence, a joint modeling framework for analyzing all longitudinal outcomes simultaneously is essential. There is a large number of joint modeling approaches for mixed type outcomes. Multivariate marginal models (e.g., likelihood-based \citep{Molenberghs2005book}, copula-based \citep{Lambert2002SIM}, and GEE-based\\ \citep{OBrien2004JRSSC}), provide direct inference for marginal treatment effects, but handling unbalanced data and more than two response variables remain open problems. Multivariate random effects models \citep{Verbeke2014SMMR} have severe computational difficulties when the number of random effects is large. In comparison, mixed effects models focused on dimensionality reduction (using latent variables) provide an excellent and balanced approach to modeling multivariate longitudinal data. To this end, \cite{He2016SMMR} developed a joint model for multiple longitudinal outcomes of mixed types, subject to an outcome-dependent terminal event. \cite{Luo_Wang2014SIM} proposed a hierarchical joint model accounting for multiple levels of correlation among multivariate longitudinal outcomes and survival data. \cite{Proust2016SIM} developed a joint model for multiple longitudinal outcomes and multiple time-to-events using shared latent classes.

In this article, we propose a novel joint model that consists of: (1) a semiparametric multilevel latent trait model (MLLTM) for the multiple longitudinal outcomes with a univariate latent variable representing the underlying disease severity, and (2) a survival submodel for the event time data. We adopt penalized splines using the truncated power series spline basis expansion in modeling the effects of some covariates and the baseline hazard function. This spline basis expansion results in tractable integration in the survival function, which significantly improves computational efficiency. We develop a Bayesian approach via Markov chain Monte Carlo (MCMC) algorithm for statistical inference and a dynamic prediction framework for the predictions of target patients' future outcome trajectories and risks of survival event. These important predictive measures offer unique insight into the dynamic nature of each patient's disease progression and they are highly relevant for patient targeting, management, prognosis, and treatment selection. Moreover, accurate prediction can advance design of future studies, experimental trials, and clinical care through improved prognosis and earlier intervention.

The rest of the article is organized as follows. In Section~\ref{sec:motivating_CT}, we describe a motivating clinical trial and the data structure. In Section~\ref{sec:methods}, we discuss the joint model, Bayesian inference, and subject-specific prediction. In Section~\ref{sec:dataAna}, we apply the proposed method to the motivating clinical trial dataset. In Section~\ref{sec:simulation}, we conduct simulation studies to assess the prediction accuracy. Concluding remarks and discussions are given in Section~\ref{sec:discussion}.

\section{A motivating clinical trial} \label{sec:motivating_CT}
The methodological development is motivated by the DATATOP study, a double-blind, placebo-controlled multicenter randomized clinical trial with 800 patients to determine if deprenyl and/or tocopherol administered to patients with early Parkinson's disease (PD) will slow the progression of PD. We refer to as placebo group the patients who did not receive deprenyl and refer to as treatment group the patients who received deprenyl. The detailed description of the design of the DATATOP study can be found in \cite{Shoulson1998AN}.

In the DATATOP study, the multiple outcomes collected include Unified PD Rating Scale (UPDRS) total score, modified Hoehn and Yahr (HY) scale, Schwab and England activities of daily living (SEADL), measured at 10 visits (baseline, month 1, and every 3 months starting from month 3 to month 24). UPDRS is the sum of 44 questions each measured on a 5-point scale (0-4), and it is approximated by a continuous variable with integer value from 0 (not affected) to 176 (most severely affected). HY is a scale describing how the symptoms of PD progresses. It is an ordinal variable with possible values at 1, 1.5, 2, 2.5, 3, 4, and 5, with higher values being clinically worse outcome. However, the DATATOP study consists of only patients with early mild PD and the worst observed HY is 3. SEADL is a measurement of activities of daily living, and it is an ordinal variable with integer values from 0 to 100 incrementing by 5, with larger values reflecting better clinical outcomes. We have recoded SEADL variable so that higher values in all outcomes correspond to worse clinical conditions and we have combined some categories with zero or small counts so that SEADL has eight categories.

\begin{figure}[h!tbp]
\centering
\includegraphics[width=0.4\textwidth,angle=270]{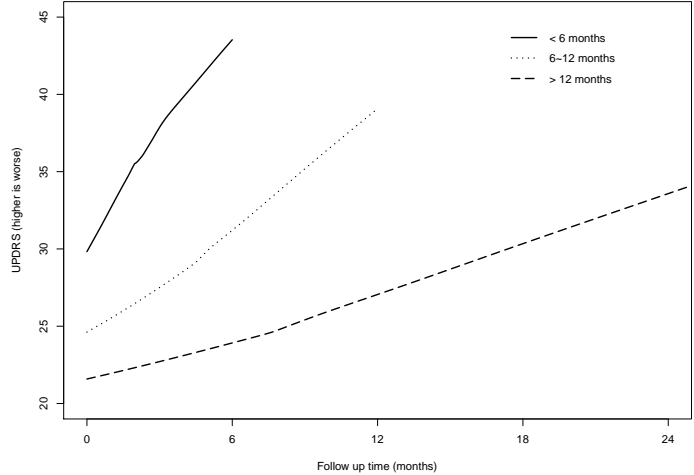}
\caption{Mean UPDRS values over time for DATATOP patients with follow-up time less than 6 months (solid line), 6-12 months (dotted line), and more than 12 months (dashed line).} \label{fig:UPDRS_fumonth}
\end{figure}

Among the 800 patients in the DATATOP study, 44 did not have disease duration recorded and one had no UPDRS measurements. We exclude them (5.6\%) from our analysis and the data analysis is based on the remaining 755 patients. The mean age of patients is 61.0 years (standard deviation, 9.5 years). 375 patients are in the placebo group and 380 are in the treatment group. About 65.8\% of patients are male and the average disease duration is 1.1 years (standard deviation, 1.1 years). Before the end of the study, some patients (207 in placebo and 146 in treatment) reached a pre-defined level of functional disability, which is considered to be a terminal event because these patients would then initiate symptomatic treatment of levodopa, which can ameliorate the clinical outcomes. Figure~\ref{fig:UPDRS_fumonth} displays the mean UPDRS measurements over time for DATATOP patients with follow-up time less than 6 months (96 patients, solid line), 6-12 months (215 patients, dotted line), and more than 12 months (444 patients, dashed line). Figure~\ref{fig:UPDRS_fumonth} suggests that patients with shorter follow-up had higher UPDRS measurements, manifesting the strong correlation between the PD symptoms and terminal event. Similar patterns are observed in HY and SEADL measurements. Such a dependent terminal event time, if not properly accounted for, may lead to biased estimates \citep{Henderson2000Biostatistics}.

Because levodopa is associated with possible motor complications \citep{Brooks2008NDT}, clinicians tend to provide more targeted interventions to delay their initiation of levodopa use. To this end, in the context of DATATOP study and similar PD studies, there is an important clinically relevant prediction question: for a new patient (not included in the DATATOP study) with one or multiple visits, what are his/her most likely future outcome trajectories (e.g., UPDRS, HY, and SEADL) and risk of functional disability within the next year, given the outcome histories and the covariate information? These important predictive measures are highly relevant for PD patient targeting, management, prognosis, and treatment selection. In this article, we propose to develop a Bayesian personalized prediction approach based on a joint modeling framework consisting of a semiparametric multilevel latent trait model (MLLTM) for multivariate longitudinal outcomes and a survival model for the event time data (time to functional disability).

\section{Methods} \label{sec:methods}
\subsection{Joint modeling framework}\label{sec:model_MLLTM}
In the context of clinical trials with multiple outcomes, the data structure is often of the type $\{y_{ik}(t_{ij}), t_i, \delta_i\}$, where $y_{ik}(t_{ij})$ is the $k$th ($k=1,\ldots,K$) outcome, which can be binary, ordinal, or continuous, for patient $i$ ($i=1,\ldots,I$) at visit $j$ ($j=1,\ldots,J_i$) recorded at time $t_{ij}$ from the study onset, $t_i=min(T_i^*,C_i)$ is the observed event time to functional disability, as the minimum between the true event time $T_i^*$ and the censoring time $C_i$ which are assumed to be independent of $T_i^*$, and $\delta_i$ is the censoring indicator ($1$ if the event is observed, and $0$ otherwise). We propose to use a semiparametric multilevel latent trait model (MLLTM) for the multiple longitudinal outcomes and a survival model for the event time.

To start building the semiparametric MLLTM framework, we assume that there is a latent variable representing the underlying disease severity score and denote it as $\theta_i(t)$ for patient $i$ at time $t$ with a higher value for more severe status. We introduce the first level model for continuous outcomes,
\begin{eqnarray}
y_{ik}(t)=a_k+b_k\theta_{i}(t)+\varepsilon_{ik}(t), \label{eqn:MLLTM_continuous}
\end{eqnarray}
where $a_k$ and $b_k$ (positive) are the outcome-specific parameters, and the random errors $\varepsilon_{ik}(t)\sim N(0, \sigma_{\varepsilon_k}^2)$. Note that $a_k=E[y_{ik}(t)|\theta_{i}(t)=0]$ is the mean of the $k$th outcome if the disease severity score is $0$ and $b_k$ is the expected increase in the $k$th outcome for one unit increase in the disease severity score. The parameter $b_k$ also plays the role of bringing up the disease severity score to the scale of the $k$th outcome. The models for outcomes that are binary (e.g., the presence of adverse events) and ordinal (e.g., HY and SEADL) are as follows \citep{Fox2005BJMSP}:
\begin{eqnarray} \label{eqn:MLLTM_ordi}
  && \textnormal{logit}\big\{p(y_{ik}(t)=1|\theta_{i}(t))\}=a_k+b_k\theta_{i}(t) \nonumber \\
  && \textnormal{logit}\big\{p(y_{ik}(t)\le l|\theta_{i}(t))\big\}=a_{kl}-b_k\theta_{i}(t),
\end{eqnarray}
where $l=1,2,\ldots,n_k-1$ is the $l$th level of the $k$th ordinal variable with $n_k$ levels. Note that the negative sign for $b_k$ in the ordinal outcome model is to ensure that worse disease severity (higher $\theta_{i}(t)$) is associated with a more severe outcome (higher $y_{ik}(t)$). Interpretation of parameters is similar for continuous outcomes, except that modeling is on the log-odds, not the native scale, of the data. We have selected logit link function in model~\eqref{eqn:MLLTM_ordi}, while other link functions (e.g., probit and complementary log-log) can be adopted. A major feature of models~\eqref{eqn:MLLTM_continuous} and \eqref{eqn:MLLTM_ordi} is that they all incorporate $\theta_{i}(t)$ and explicitly combine longitudinal information from all outcomes.

To model the dependence of severity score $\theta_{i}(t)$ on covariates, we propose the second level semiparametric model
\begin{eqnarray}\label{eqn:theta}
\theta_{i}(t)=\boldsymbol{X}_{i}(t)\boldsymbol{\beta} + \boldsymbol{Z}_{i}(t)\boldsymbol{u}_i + \boldsymbol{V}_R(t) \boldsymbol{\zeta},
\end{eqnarray}
where vectors $\boldsymbol{X}_{i}(t)$ and $\boldsymbol{Z}_{i}(t)$ are $p$ and $q$ dimensional covariates corresponding to fixed and random effects, respectively. They can include covariates of interest such as treatment and time. To allow additional flexibility and smoothness in modeling the effects of some covariates, we adopt a smooth time function $\boldsymbol{V}_R(t) \boldsymbol{\zeta}=\sum_{r=1}^{R}\zeta_{r}(t-\kappa_r)_+$ using the truncated power series spline basis expansion $\boldsymbol{V}_R(t) = \{ (t-\kappa_1)_+, \ldots, (t-\kappa_R)_+ \}$, where $\boldsymbol{\kappa} = \{\kappa_1, \ldots, \kappa_R \}$ are the knots, and $(t-\kappa_r)_+ = t-\kappa_r$ if $t>\kappa_r$ and 0 otherwise. Following \cite{Ruppert2012JCGS}, we consider a large number of knots (typically 5 to 20) that can ensure the desired flexibility and we select the knot location to have sufficient subjects between adjacent knots. To avoid overfitting, we explicitly introduce smoothing by assuming that $\boldsymbol{\zeta} = (\zeta_1, \ldots, \zeta_R)' \sim N(0, \sigma_\zeta^2 \boldsymbol{I})$\citep{Ruppert2003semipar_book,Crainiceanu2005JSS}. The choice of knots is important to obtain a well fitted model and should be selected with caution to avoid overfitting. Several approaches of automatic knot selection based on stepwise model selection have been proposed \citep{Friedman1989Technometrics,Stone1997AOS,Denison1998JRSSB,Dimatteo2001Biometrika}. \cite{Wand2000CS} gives a good review and comparison of some of these approaches. Penalizing the spline coefficients to constrain their influence also helps to avoid overfitting \citep{Ruppert2003semipar_book}, as in our model. Moreover, in clinical studies with same scheduled follow-up visits, the frequency of study visits needs to be accounted for in the selection of knots. For the ease of illustration, we include the nonparametric smooth function for the time variable, although our model can be extended to accommodate more nonparametric smooth functions. The vector $\boldsymbol{u}_i=(u_{i1}, \ldots, u_{iq})'$ contains the random effects for patient $i$'s latent disease severity score and it is distributed as $N(\boldsymbol{0}, \boldsymbol{\Sigma})$. Equations~\eqref{eqn:MLLTM_continuous}, \eqref{eqn:MLLTM_ordi} and \eqref{eqn:theta} consist of the semiparametric MLLTM model, which provides a nature framework for defining the overall effects of treatment and other covariates. Indeed, if $\theta_{i}(t)=\beta_{0}+\beta_{1}x_i+\beta_{2}t+\beta_{3}x_i t + \sum_{r=1}^{R} \zeta_r (t-\kappa_r)_+ +u_{i0}+u_{i1}t$, where $x_i$ is treatment indicator (1 if treatment and 0 otherwise), then $\beta_1$ is the main treatment effect and $\beta_3$ is the time-dependent treatment effect. In this context, the null hypothesis of no overall treatment effect is $H_0: \beta_1=\beta_3=0$. Because the number of outcomes ($K$) has been reduced to one latent disease severity score, models are quite parsimonious in terms of number of random effects, which improves computational feasibility and model interpretability.

Because the semiparametric MLLTM model is over-parameterized, additional constraints are required to make it identifiable. Specifically, we set $a_{k1}=0$ and $b_k=1$ for one ordinal outcome. For the ordinal outcome $k$ with $n_k$ categories, the order constraint $a_{k1} < \ldots < a_{kl} < \ldots < a_{kn_{k}-1}$ must be satisfied, and the probability of being in a particular category is $p(Y_{ik}(t)=l)= p(Y_{ik}(t)\leq l|\theta_{i}(t))- p(Y_{ik}(t)\leq l-1|\theta_{i}(t))$. With these assumptions, the conditional log-likelihood of observing the patient $i$ data $\{y_{ik}(t_{ij})\}$ given $\boldsymbol{u}_i$ and $\boldsymbol{\zeta}$ is $l_y(\boldsymbol{\Theta}_y;\boldsymbol{y}_i,\boldsymbol{u}_i, \boldsymbol{\zeta})=\sum^{J_i}_{j=1}\sum^K_{k=1} \log p(y_{ik}(t_{ij})|\boldsymbol{u}_i, \boldsymbol{\zeta})$. For notational convenience, we let $\boldsymbol{a}=(\boldsymbol{a}'_1,\ldots,\boldsymbol{a}'_k,\ldots,\boldsymbol{a}'_K)'$, with $\boldsymbol{a}_k$ being numeric for binary and continuous outcomes and $\boldsymbol{a}_k=(a_{k1},\ldots,a_{kn_{k}-1})'$ for ordinal outcomes. We let $\boldsymbol{b}=(b_1,\ldots,b_K)'$ and $\boldsymbol{y}_{i}(t)=\{y_{ik}(t), k=1, \ldots, K\}'$ be the vector of measurements for patient $i$ at time $t$ and let $\boldsymbol{y}_i=\{\boldsymbol{y}_{i}(t_{ij}), j=1, \ldots, J_i\}$ be the outcome vector across $J_i$ visit times. The parameter vector for the longitudinal process is $\boldsymbol{\Theta}_y = (\boldsymbol{a}', \boldsymbol{b}', \boldsymbol{\beta}', \boldsymbol{\Sigma}, \sigma_{\varepsilon_k}, \sigma_\zeta)'$.

To model the survival process, we use the proportional hazard model
\begin{equation}\label{eqn:Cox}
h_i(t)=h_0(t)\exp\{\boldsymbol{W}_i\boldsymbol{\gamma}+\nu \theta_i(t)\},
\end{equation}
where $\boldsymbol{\gamma}$ is the coefficient for time-independent covariates $\boldsymbol{W}_i$ and $h_0(\cdot)$ is the baseline hazard function. Some covariates in $\boldsymbol{W}_i$ can overlap with vector $\boldsymbol{X}_i(t)$ in model~\eqref{eqn:theta}. \cite{Ibrahim2010JCO} gave an excellent explanation of the coefficients for those overlapped covariates. In the current context, if we denote $\boldsymbol{\beta}_o$ and $\boldsymbol{\gamma}_o$ as the coefficients for the overlapped covariates in vectors $\boldsymbol{X}_i(t)$ and $\boldsymbol{W}_i$, respectively, we have: (1) $\boldsymbol{\beta}_o$ is the covariate effect on the longitudinal latent variable; (2) $\boldsymbol{\gamma}_o$ is the direct covariate effect on the time to event; (3) $\nu\boldsymbol{\beta}_o+\boldsymbol{\gamma}_o$ is the overall covariate effect on the time to event. The association parameter $\nu$ quantifies the strength of correlation between the latent variable $\theta_i(t)$ and the hazard for a terminal event at the same time point (refer to as \lq Model 1: shared latent variable model\rq). Specifically, a value of $\nu=0$ indicates that there is no association between the latent variable and the event time while a positive association parameter $\nu$ implies that patients with worse disease severity tend to have a terminal event earlier, e.g., a value of $\nu=0.5$ indicates that the hazard rate of having the terminal event increases by $65\%$ (i.e., $\exp(0.5)-1$) for every unit increase in the latent variable. For prediction of subject-specific survival probabilities, a specified and smooth baseline hazard function is desired. To this end, we again adopt a truncated power series spline basis expansion $h_0(t) = \exp\{\eta_{0}+\eta_{1}t + \sum_{r=1}^{R}\xi_{r}(t-\kappa_r)_+\}$ and assume $\boldsymbol{\xi} = (\xi_1, \ldots, \xi_R )' \sim N(0, \sigma_{\xi}^2\boldsymbol{I})$ to introduce smoothing. The knot locations can be the same or different from those in equation~\eqref{eqn:theta}.

In equation~\eqref{eqn:Cox}, different formulations can be used to postulate how the risk for a terminal event depends on the unobserved disease severity score at time $t$. For example, one can add to equation~\eqref{eqn:Cox} a time-dependent slope $\theta_i'(t)$, so that the risk depends on both the current severity score and the slope of the severity trajectory at time $t$ (refer to as \lq Model 2: time-dependent slope model\rq):
\begin{equation}
h_i(t)=h_0(t)\exp\{\boldsymbol{W}_i\boldsymbol{\gamma}+\nu_1 \theta_i(t)+\nu_2 \theta_i'(t)\}.
\end{equation}
Alternatively, one can consider the standard formulations of joint models that include only the random effects in the Cox model (refer to as \lq Model 3: shared random effects model\rq):
\begin{equation}
h_i(t)=h_0(t)\exp\{\boldsymbol{W}_i\boldsymbol{\gamma}+ \boldsymbol{\nu}' \boldsymbol{u}_i \}.
\end{equation}
A good summary of these various formulations in the joint modeling framework can be found in \cite{Rizopoulos2014JASA} and \cite{Yang2015SIM}.

The log-likelihood of observing event outcome $t_i$ and $\delta_i$ for patient $i$ is \\ $l_s(\boldsymbol{\Theta}_s; t_i, \delta_i, \boldsymbol{u}_i, \boldsymbol{\zeta}, \boldsymbol{\xi})=\log \{h_i(t_i)^{\delta_i}S_i(t_i)\}$, where the survival function $S_i(t_i)=\exp\{-\int_0^{t_i} h_i(s)ds\}$ and the parameter vector for the survival process is $\boldsymbol{\Theta}_s = (\boldsymbol{\gamma}', \nu, \eta_0, \eta_1, \sigma_\xi)'$. Note that the truncated power series spline basis expansion in modeling the smooth time function in equation~\eqref{eqn:theta} and in modeling the baseline hazard function is linear function of time, which results in tractable integration in the survival function $S_i(t_i)$, and consequently,  significant gain in computing efficiency. Conditional on the random effect vector $\boldsymbol{u}_i$, $\boldsymbol{y}_i$ is assumed to be independent of $t_i$. The penalized log-likelihood of the joint model for patient $i$ given random effects $\boldsymbol{u}_i$ and smoothing parameters $\sigma_\zeta$, $\sigma_\xi$ is
\begin{equation}\label{eqn:joint_Lik}
l(\boldsymbol{\Theta}, \boldsymbol{\zeta}, \boldsymbol{\xi};\cdot)=l_y(\boldsymbol{\Theta}_y;\boldsymbol{y}_i,\boldsymbol{u}_i, \boldsymbol{\zeta}) + l_s(\boldsymbol{\Theta}_s;t_i,\delta_i, \boldsymbol{u}_i, \boldsymbol{\zeta}, \boldsymbol{\xi}) - \frac{1}{\sigma_\zeta^2} \boldsymbol{\zeta}'\boldsymbol{\zeta} - \frac{1}{\sigma_\xi^2} \boldsymbol{\xi}'\boldsymbol{\xi},
\end{equation}
where the unknown parameter vector $\boldsymbol{\Theta} = (\boldsymbol{\Theta}'_y, \boldsymbol{\Theta}'_s)'$.

\subsection{Bayesian inference} \label{sec:BayesianInf}
To infer the unknown parameter vector $\boldsymbol{\Theta}$, we use Bayesian inference based on Markov chain Monte Carlo (MCMC) posterior simulations. The fully Bayesian inference has many advantages. First, MCMC algorithms can be used to estimate exact posterior distributions of the parameters, while likelihood-based estimation only produces a point estimate of the parameters, with asymptotic standard errors \citep{Dunson2007SMMR}. Second, Bayesian inference provides better performance in small samples compared to likelihood-based estimation \citep{Lee2004MBR}. In addition, it is more straightforward to deal with more complicated models using Bayesian inference via MCMC. We use vague priors on all elements in $\boldsymbol{\Theta}$. Specifically, the prior distributions of parameters $\nu$, $\eta_0$, $\eta_1$, and all elements in vectors $\boldsymbol{\beta}$ and $\boldsymbol{\gamma}$ are $N(0,100)$. We use the prior distribution $b_k\sim\textnormal{Uniform}(0, 10)$, $k=2, \ldots, K$, to ensure positivity. The prior distribution for the difficulty parameter $a_k$ of the continuous outcomes is $a_k\sim N(0, 100)$. To obtain the prior distributions for the threshold parameters of ordinal outcome $k$, we let $a_{k1}\sim N(0, 100)$, and $a_{kl}=a_{k,l-1}+\Delta_l$ for $l=2, \ldots, n_k-1$, with $\Delta_l\sim N(0, 100)I(0,)$, i.e., normal distribution left truncated at $0$. We use the prior distribution $\textnormal{Uniform}[-1,1]$ for all the correlation coefficients $\rho$ in the covariance matrix $\boldsymbol{\Sigma}$, and $\textnormal{Inverse-Gamma}(0.01, 0.01)$ for all variance parameters. We have investigated other selections of vague prior distributions with various hyper-parameters and obtained very similar results.

The posterior samples are obtained from the full conditional of each unknown parameter using Hamiltonian Monte Carlo (HMC) \citep{Duane1987Physics_Letters} and No-U-Turn Sampler (NUTS, a variant of HMC) \citep{hoffman-gelman:2013}. Compared with the Metropolis-Hastings algorithm, HMC and NUTS reduce the correlation between successive sampled states by using a Hamiltonian evolution between states and by targeting states with a higher acceptance criteria than the observed probability distribution, leading to faster convergence to the target distribution. Both HMC and NUTS samplers are implemented in \cp{Stan}, which is a probabilistic programming language implementing statistical inference. The model fitting is performed in \cp{Stan} (version $2.14.0$) \citep{stan-manual:2016} by specifying the full likelihood function and the prior distributions of all unknown parameters. For large dataset, \cp{Stan} may be more efficient than \cp{BUGS} language \citep{Lunn2000Winbugs} in achieving faster convergence and requiring smaller number of samples \citep{hoffman-gelman:2013}. To monitor Markov chain convergence, we use the history plots and view the absence of apparent trends in the plot as evidence of convergence. In addition, we use the Gelman-Rubin diagnostic to ensure the scale reduction $\widehat{R}$ of all parameters are smaller than $1.1$ as well as a suite of convergence diagnosis criteria to ensure convergence \citep{Gelman2013BDA3}. After fitting the model to the training dataset (the dataset used to build the model) using Bayesian approaches via MCMC, we obtain $M$ (e.g., $M=2,000$ after burn-in) samples for the parameter vector $\boldsymbol{\Theta}_0 = (\boldsymbol{\Theta}', \boldsymbol{\zeta}', \boldsymbol{\xi}')'$. To facilitate easy reading and implementation of the proposed joint model, a \cp{Stan} code has been posted in the \ref{sec:WebSupp}. Note that \cp{Stan} requires variable types to be declared prior to modeling. The declaration of matrix $\boldsymbol{\Sigma}$ as a covariance matrix ensures it to be positive-definite by rejecting the samples that cannot produce positive-definite matrix $\boldsymbol{\Sigma}$. Please refer to the \cp{Stan} code in the \ref{sec:WebSupp} for details.

\subsection{Dynamic prediction framework}\label{sec:IndPred}
We illustrate how to make prediction for a new subject $N$, based on the available outcome histories $\boldsymbol{y}^{\{t\}}_N=\{\boldsymbol{y}_N(t_{Nj}); 0 \le t_{Nj} \le t \}$ and the covariate history $\boldsymbol{X}^{\{t\}}_N=\{\boldsymbol{X}_N(t_{Nj}), \boldsymbol{Z}_N(t_{Nj}),$ $\boldsymbol{W}_N; 0 \le t_{Nj} \le t \}$ up to time $t$, and $\delta_N=0$ (no event). We want to obtain two personalized predictive measures: the longitudinal trajectories $y_{Nk}(t')$, for $k=1, \ldots, K$, at a future time point $t'>t$ (e.g., $t'=t+\Delta t$), and the probability of functional disability before time $t'$, denoted by $\pi_N(t'|t)=p(T^*_N \le t'|T^*_N>t, \boldsymbol{y}^{\{t\}}_N, \boldsymbol{X}^{\{t\}}_N)$. To do this, the key step is to obtain samples for patient $N$'s random effects vector $\boldsymbol{u}_N$ from its posterior distribution $p(\boldsymbol{u}_N|T_N^*>t, \boldsymbol{y}^{\{t\}}_N, \boldsymbol{\Theta}_0)$. Specifically, conditional on the $m$th posterior sample $\boldsymbol{\Theta}_0^{(m)}$, we draw the $m$th sample of the random effects vector $\boldsymbol{u}_N$ from its posterior distribution
\begin{eqnarray*}
p(\boldsymbol{u}_N|T_N^*>t, \boldsymbol{y}^{\{t\}}_N, \boldsymbol{\Theta}_0^{(m)}) &=& \frac{p(\boldsymbol{y}^{\{t\}}_N,T_N^*>t, \boldsymbol{u}_N|\boldsymbol{\Theta}_0^{(m)})}{p(\boldsymbol{y}^{\{t\}}_N,T_N^*>t|\boldsymbol{\Theta}_0^{(m)})} \propto p(\boldsymbol{y}^{\{t\}}_N, T_N^*>t,\boldsymbol{u}_N|\boldsymbol{\Theta}_0^{(m)}) \nonumber \\
&=& p(\boldsymbol{y}^{\{t\}}_N | \boldsymbol{u}_N, \boldsymbol{\Theta}_0^{(m)}) p(T_N^*>t|\boldsymbol{u}_N, \boldsymbol{\Theta}_0^{(m)}) p(\boldsymbol{u}_N|\boldsymbol{\Theta}_0^{(m)}),
\end{eqnarray*}
where the first equality is from Bayes theorem.

For each of $\boldsymbol{\Theta}_0^{(m)}$, $m=1, \ldots, M$, we use adaptive rejection Metropolis sampling \citep{Gilks1995AppStat} to draw 50 samples of random effects vector $\boldsymbol{u}_N$ and retain the final sample. This process is repeated for the $M$ saved values of $\boldsymbol{\Theta}_0$. Suppose that patient $N$ does not develop functional disability by time $t'$, then the outcome histories are updated to $\boldsymbol{y}^{\{t'\}}_N$. We can dynamically update the posterior distribution to $p(\boldsymbol{u}_N|T_N^*>t', \boldsymbol{y}^{\{t'\}}_N, \boldsymbol{\Theta}_0^{(m)})$, draw new samples, and obtain the updated predictions.

With the $M$ samples for patient $N$'s random effects vector $\boldsymbol{u}_N$, predictions can be obtained by simply plugging in realizations of the parameter vector and random effects vector $\{\boldsymbol{\Theta}_0^{(m)}, \boldsymbol{u}_N^{(m)}, m=1,\ldots,M\}$. For example, the $m$th sample of continuous outcome $y_{Nk}(t')$ is obtained from equations~\eqref{eqn:MLLTM_continuous} and \eqref{eqn:theta}:
\begin{equation*}
y_{Nk}^{(m)}(t')=a_k^{(m)}+b_k^{(m)}\left\{ \boldsymbol{X}_N(t')\boldsymbol{\beta}^{(m)} + \boldsymbol{Z}_N(t')\boldsymbol{u}_N^{(m)} + \boldsymbol{V}_R(t')\boldsymbol{\zeta}^{(m)} \right\} +\varepsilon_{Nk}^{(m)}(t'),
\end{equation*}
where the random errors $\varepsilon_{Nk}^{(m)}(t') \sim N (0, \sigma_{\varepsilon_k}^{2(m)})$, and each parameter is replaced by the corresponding element in the $m$th sample $\{\boldsymbol{\Theta}_0^{(m)}, \boldsymbol{u}_N^{(m)}\}$.

Similarly, the $m$th sample of ordinal outcome $y_{Nk}(t')=l$ with $l=1, 2, \ldots, n_k$ is
\begin{equation*}
\textnormal{logit}\big\{p\big(y_{Nk}^{(m)}(t') \le l\big)\big\}=a_{kl}^{(m)} - b_k^{(m)}\big\{ \boldsymbol{X}_N(t')\boldsymbol{\beta}^{(m)} + \boldsymbol{Z}_N(t')\boldsymbol{u}_N^{(m)} + \boldsymbol{V}_R(t')\boldsymbol{\zeta}^{(m)} \big\}.
\end{equation*}
The probability of being in category $l$ is $p\big(y_{ik}^{(m)}(t')=l\big) = p\big(y_{ik}^{(m)}(t') \le l\big) - p\big(y_{ik}^{(m)}(t') \le l-1\big)$. The $m$th sample of the hazard of patient $i$ at time $t'$ is
\begin{equation*}
h_N^{(m)}(t'|\boldsymbol{u}_N^{(m)}) = h_0^{(m)}(t')\exp\big\{\boldsymbol{W}_N\boldsymbol{\gamma}^{(m)} + \nu^{(m)}\big[\boldsymbol{X}_N(t')\boldsymbol{\beta}^{(m)} + \boldsymbol{Z}_N(t')\boldsymbol{u}_N^{(m)} + \boldsymbol{V}_R(t')\boldsymbol{\zeta}^{(m)}\big] \big\}.
\end{equation*}
Thus, the conditional probability of functional disability before time $t'$ is
\begin{eqnarray*}
\widehat{\pi}_N(t'|t) &=& \int p(T^*_N \le t'|T^*_N>t, \boldsymbol{y}^{\{t\}}_N, \boldsymbol{X}^{\{t\}}_N, \boldsymbol{u}_N) p(\boldsymbol{u}_N|T^*_N>t, \boldsymbol{y}^{\{t\}}_N, \boldsymbol{X}^{\{t\}}_N )d\boldsymbol{u}_N \nonumber \\
&\approx & \frac{1}{M}\sum_{m=1}^M p\left(T_N^* \le t'|T_N^*>t, \boldsymbol{y}^{\{t\}}_N, \boldsymbol{X}^{\{t\}}_N, \boldsymbol{u}_N^{(m)}\right) \nonumber \\
&=& \frac{1}{M}\sum_{m=1}^M \left\{1-\frac{p(T_N^* > t'|\boldsymbol{y}^{\{t\}}_N, \boldsymbol{X}^{\{t\}}_N, \boldsymbol{u}_N^{(m)})} {p(T_N^* > t|\boldsymbol{y}^{\{t\}}_N, \boldsymbol{X}^{\{t\}}_N, \boldsymbol{u}_N^{(m)})} \right\} \\
&=& \frac{1}{M}\sum_{m=1}^M \left\{1 - \exp\left(-\int_t^{t'}h_N^{(m)}(s|\boldsymbol{u}_N^{(m)})ds\right) \right\},
\end{eqnarray*}
where the integration with respect to $\boldsymbol{u}_N$ in the first equality is approximated using Monte Carlo method. Note that the truncated power series spline basis expansion in modeling the smooth time function in equation~\eqref{eqn:theta} and in modeling the baseline hazard function results in tractable integration not only in the survival function $S_N(t_N)$, but also in the integration of hazard function in the last equality. All prediction results can then be obtained by calculating simple summaries (e.g., mean, variance, quantiles) of the posterior distributions of $M$ samples $\big\{y_{Nk}^{(m)}(t'), m=1,\ldots,M \big\}$. Note that although it may take a few hours to obtain enough posterior samples for the parameter vector $\boldsymbol{\Theta}_0$, it only takes a few seconds to obtain the prediction results for a new subject. Hence, the dynamic prediction framework and the web-based calculator (detailed in Section~4) can provide instantaneous supplemental information for PD clinicians to monitor disease progression.

\subsection{Assessing predictive performance}
It is essential to assess the performance of the proposed predictive measures. Here, we focus on the probability $\pi(t'|t)$. Specifically, we assess the discrimination (how well the models discriminate between patients who had the event from patients who did not) using the receiver operating characteristic (ROC) curve and the area under the ROC curves (AUC) and assess the validation (how well the models predict the observed data) using the expected Brier score (BS).

\subsubsection{Area under the ROC curves}
Following the notation in Section~\ref{sec:IndPred}, for any given cut point $c\in(0,1)$, the time-dependent sensitivity and specificity are defined as $\textnormal{sensitivity}(c,t,t'):P\left\{\pi_i(t'|t) > c|N_i(t,t')=1, T_i^*>t\right\}$ and $\textnormal{specificity}(c,t,t'):P\left\{\pi_i(t'|t) \le c|N_i(t,t')=0, T_i^*>t\right\}$, respectively, where $N_i(t,t')=I(t < T_i^* \le t')$, indicating whether there is an event (case) or no event (control) observed for subject $i$ during the time interval $(t, t']$. In the absence of censoring, sensitivity and specificity can be simply estimated from the empirical distribution of the predicted risk among either cases or controls. To handle censored event times, \cite{Li2016SMMR} proposed an estimator for the sensitivity and specificity based on the predictive distribution of the censored survival time:
 \begin{eqnarray} \label{eqn:simpleROC}
\widehat{P}\left\{\pi_i(t'|t) > c|N_i(t,t')=1, T_i^*>t\right\} = \frac{\sum_{i=1}^{n}\widehat{W}_i(t,t') I\{\widehat{\pi}_i(t'|t) > c \}}{\sum_{i=1}^{n}\widehat{W}_i(t,t')}  \\
\widehat{P}\left\{\pi_i(t'|t) \le c|N_i(t,t')=0, T_i^*>t\right\} = \frac{\sum_{i=1}^{n}[1-\widehat{W}_i(t,t')] I\{\widehat{\pi}_i(t'|t) \le c \}}{\sum_{i=1}^{n}[1-\widehat{W}_i(t,t')]}, \nonumber
\end{eqnarray}
where $\widehat{W}_i(t, t')$ is the weight to account for censoring and it is defined as
\begin{eqnarray*}
\widehat{W}_i(t, t') &=& I(t<t_i \le t')\delta_i + I(t<t_i \le t')(1-\delta_i)P\{T_i^* < t'|T_i^* \ge t_i, \widehat{\pi}_i(t'|t)\} \nonumber \\
&=& I(t<t_i \le t')\delta_i + I(t<t_i \le t')(1-\delta_i)\left[1-\frac{P\{T_i^* \ge t'|\widehat{\pi}_i(t'|t)\}}{P\{T_i^* \ge t_i|\widehat{\pi}_i(t'|t)\}} \right].
\end{eqnarray*}
Note that the subjects who have the survival event before time $t$ (i.e., $t_i < t$) have their estimated weight $\widehat{W}_i(t, t')=0$ and thus they play no role in equation~\eqref{eqn:simpleROC}. The conditional survival distribution $P\{T_i^* \ge \tilde{t}|\widehat{\pi}_i(t'|t)\}$, where $\tilde{t}$ can be either $t'$ or $t_i$, can be estimated using kernel weighted Kaplan-Meier method with a bandwidth $d$, which can be easily implemented in standard survival analysis software accommodating weighted data:
\begin{equation*}
P\{T_i^* \ge \tilde{t}|\widehat{\pi}_i(t'|t)\} = \prod_{s \in \Omega, s \le \tilde{t}} \left[1 - \frac{\sum_{i' \neq i}K_d\{\widehat{\pi}_{i'}(t'|t), \widehat{\pi}_i(t'|t)\}I(T_{i'}=s)\delta_{i'}} {\sum_{i' \neq i}K_d\{\widehat{\pi}_{i'}(t'|t), \widehat{\pi}_i(t'|t)\}I(T_{i'}\ge s)}   \right],
\end{equation*}
where $\Omega$ is the set of distinct $t_i$'s with $\delta_i=1$ and $K_d$ is the kernel function, e.g., uniform and Gaussian kernels. Specifically, we use uniform kernel in this article.

With the estimation of sensitivity and specificity, the time-dependent ROC curve can be constructed for all possible cut points $c\in(0, 1)$ and the corresponding time-dependent $\textnormal{AUC}(t, t')$ can be estimated using standard numerical integration methods such as Simpson's rule.

\subsubsection{Dynamic Brier score} The Brier score (BS) developed in survival models can be extended to joint models for prediction validation \citep{Sene2016SMMR,Proust2014SMMR}. The dynamic expected BS is defined as $E[(D(t'|t) -\pi(t'|t))^2]$, where the observed failure status $D(t'|t)$ equals to 1 if the subject experiences the terminal event within the time interval $(t, t']$ and 0 if the subject is event free until $t'$. An estimator of BS is
\begin{equation*}
\widehat{\textnormal{BS}}(t, t') = \frac{1}{N_t} \sum_{i=1}^{N_t} \widehat{G}_i(t, t')\left(D_i(t, t')-\pi_i(t'|t)\right)^2,
\end{equation*}
where $N_t$ is the number of subjects at risk at time $t$, and the weight $\widehat{G}_i(t, t')= \frac{I(t_i>t')}{\hat{S}_0(t')/\hat{S}_0(t)} + \frac{I(t<t_i \le t')\delta_i}{\hat{S}_0(t_i)/\hat{S}_0(t)}$ is to account for censoring with $\hat{S}_0$ denoting the Kaplan-Meier estimate \citep{Sene2016SMMR}.

AUC and BS complement each other by assessing different aspects of the prediction. AUC has a simple interpretation as a concordance index, while BS accounts for the bias between the predicted and true risks. In general, $\textnormal{AUC}=1$ indicates perfect discrimination and $\textnormal{AUC}=0.5$ means no better than random guess, while $\textnormal{BS} = 0$ indicates perfect prediction and $\textnormal{BS} = 0.25$ means no better than random guess. \cite{Blanche2015Biometrics} provides excellent illustration of AUC and BS.

\section{Application to the DATATOP study}\label{sec:dataAna}
In this section, we apply the proposed joint model and prediction process to the motivating DATATOP study. For all results in this section, we run two parallel MCMC chains with overdispersed initial values and run each chain for $2,000$ iterations. The first $1,000$ iterations are discarded as burn-in and the inference is based on the remaining $1,000$ iterations from each chain. Good mixing properties of the MCMC chains for all model parameters are observed in the trace plots. The scale reduction $\widehat{R}$ of all parameters are smaller than $1.1$.

In order to validate the prediction and compare the performance of candidate models, we conduct a 5-fold cross-validation, where 4 partitions of the data are used to train the model and the left-out partition is used for validation and model selection. Then we fit the final selected model to the whole dataset, except that 2 patients are set aside for subject-specific prediction purpose. The covariates of interest included in equation~\eqref{eqn:theta} are baseline disease duration, baseline age, treatment (active deprenyl only), time, and the interaction term of treatment and time. We allow a flexible and smooth disease progression along time by using penalized truncated power series splines with 7 knots at the location $\boldsymbol{\kappa} = (1.2, 3, 6, 9, 12, 15, 18)$ in months, to ensure sufficient patients within each interval. Specifically, euqation~\eqref{eqn:theta} is
\begin{eqnarray*}
\theta_i(t_{ij}) &=& \beta_{0} + \beta_{1}\textnormal{duration}_i + \beta_{2}\textnormal{age}_i + \beta_{3}\textnormal{trt}_i + \beta_{4}t_{ij} \nonumber \\
&& + \beta_{5}(\textnormal{trt}_i \times t_{ij}) + \sum_{r=1}^{7}\zeta_{r}(t_{ij}-\kappa_{r})_+ + u_{i0} + u_{i1}t_{ij},
\end{eqnarray*}
where the random effects $(u_{i0}, u_{i1})' \sim N_2(0, \boldsymbol{\Sigma})$ with $\boldsymbol{\Sigma} = \{(\sigma_1^2, \rho\sigma_1\sigma_2), (\rho\sigma_1\sigma_2, \sigma_2^2)\}$ and $\boldsymbol{\zeta} \sim N(0, \sigma_{\zeta}^2\boldsymbol{I})$ to avoid overfitting.

For the survival part, three different formulations are considered as discussed in Section \ref{sec:model_MLLTM}. For instance, the shared latent variable model (Model 1) is $h_i(t) = h_0(t)\exp (\gamma_1\textnormal{duration}_i + \gamma_2\textnormal{age}_i + \gamma_3\textnormal{trt}_i + \nu \theta_i(t))$. The proposed time-dependent slope model (Model 2) and shared random effects model (Model 3) can be obtained by replacing $\nu \theta_i(t)$ with $\nu_1 \theta_i(t)+\nu_2 \theta_i'(t)$ and $\boldsymbol{\nu}' \boldsymbol{u}_i$, respectively. The baseline hazard $h_0(t)$ is similarly approximated by penalized splines $h_0(t) = \exp\{\eta_{0}+\eta_{1}t + \sum_{r=1}^{7}\xi_{r}(t-\kappa_r)_+\}$ and $\boldsymbol{\xi} \sim N(0, \sigma_{\xi}^2\boldsymbol{I})$. In addition, we compared the proposed model with two standard predictive models for time to event data, (1) a widely used univariate joint model (refer to as Model JM), where the continuous UPDRS is used as the longitudinal outcome regressing on same covariates of interest and the survival part is constructed in the same structure, and (2) a naive Cox model adjusted for time-independent covariates including all baseline characteristics as well as UPDRS, HY and SEADL scores.

We compare the performance of all candidate models in terms of discrimination and validation using 5-fold cross-validation and present AUC and BS score in Table \ref{tab:AUC_BS} and Web Table \ref{tab:S1}. All of the three formulations of the proposed MLLTM joint model outperform the univariate Model JM (except $\textnormal{AUC}(t=3, t'=9)$) and naive Cox model with larger AUC and smaller BS in most of the scenarios, suggesting that the MLLTM model accounting for multivariate longitudinal outcomes are preferable in terms of prediction. The three formulations have very similar performance with close AUC and BS. Model 1 is selected as our final model, because it leads to a straightforward interpretation of the overall covariate effect described in Section \ref{sec:model_MLLTM} and it is more intuitive to use the trajectory of latent variable $\theta_i(t)$ to predict the time to event as in Model 1, instead of using time-dependent slope $\theta'_i(t)$ or random effects $\boldsymbol{u}_i$ as in Models 2 and 3. The results also suggest that AUC increases by using more follow up measurements, e.g., in Model 1, conditional on the the measurement history up to month 3 (i.e., $t=3$), when $t'=15$, $\textnormal{AUC}(t=3, t'=15)=0.744$, while AUC increase to $\textnormal{AUC}(t=12, t'=15)=0.766$, indicating that conditional on the measurement history up to month 12, our model has 0.766 probabilities to correctly assign higher probability of functional disability by month 15 to more severe patients (who had functional disability earlier) than less severe patients (who had functional disability later). Meanwhile, BS decreases from $\textnormal{BS}(3,15)=0.216$ to $\textnormal{BS}(12,15) = 0.108$, i.e., the mean square error of prediction decreases from 0.216 to 0.108, suggesting better prediction in terms of validation.

\begin{table}[ht]
\centering
\caption{Area under the ROC curve and Brier score (BS) for the DATATOP study.} \label{tab:AUC_BS}
\begin{tabular}{rrcccccccccc}
  \hline
    \multirow{2}{*}{$t$} & \multirow{2}{*}{$t'$} & \multicolumn{2}{c}{Model 1} & \multicolumn{2}{c}{Model 2} & \multicolumn{2}{c}{Model 3} & \multicolumn{2}{c}{Model JM} & \multicolumn{2}{c}{Cox} \\
  \cline{3-12}
   & & AUC & BS & AUC & BS & AUC & BS & AUC & BS & AUC & BS  \\
  \hline
  3 & 9 & 0.754 & 0.136 & 0.759 & 0.138 & 0.761 & 0.139 & 0.757 & 0.140 & 0.736 & 0.139 \\
    & 12 & 0.744 & 0.204 & 0.744 & 0.200 & 0.744 & 0.200 & 0.739 & 0.203 & 0.725 & 0.203 \\
    & 15 & 0.744 & 0.216 & 0.742 & 0.212 & 0.744 & 0.211 & 0.726 & 0.218 & 0.719 & 0.212 \\
    & 18 & 0.775 & 0.171 & 0.766 & 0.163 & 0.772 & 0.167 & 0.728 & 0.186 & 0.720 & 0.185 \\
	\\
  6 & 9 & 0.789 & 0.078 & 0.806 & 0.078 & 0.806 & 0.078 & 0.770 & 0.081 & 0.721 & 0.094 \\
    & 12 & 0.764 & 0.159 & 0.778 & 0.154 & 0.775 & 0.154 & 0.732 & 0.164 & 0.705 & 0.173 \\
    & 15 & 0.763 & 0.183 & 0.771 & 0.178 & 0.771 & 0.178 & 0.725 & 0.194 & 0.697 & 0.194 \\
    & 18 & 0.786 & 0.158 & 0.773 & 0.154 & 0.769 & 0.159 & 0.726 & 0.175 & 0.701 & 0.175 \\
	\\
  12 & 15 & 0.766 & 0.108 & 0.787 & 0.103 & 0.782 & 0.102 & 0.695 & 0.124 & 0.647 & 0.155 \\
     & 18 & 0.758 & 0.149 & 0.739 & 0.147 & 0.723 & 0.153 & 0.700 & 0.161 & 0.663 & 0.163 \\
   \hline
\end{tabular}
\end{table}

\begin{figure}[h!tbp]
\centering
\includegraphics[width=1\textwidth,angle=0]{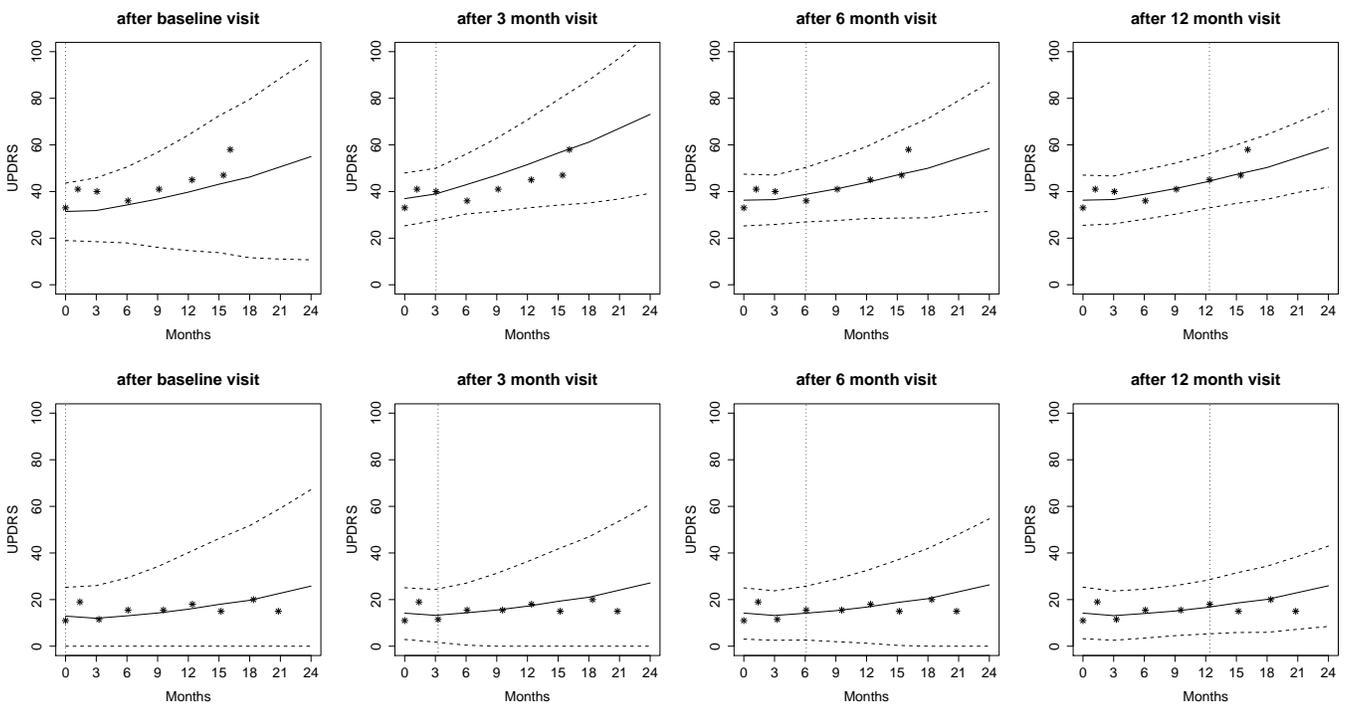}
\caption{Predicted UPDRS for Patient 169 (upper panels) and Patient 718 (lower panels). Solid line is the mean of 2000 MCMC samples. Dashed lines are the 2.5\% and 97.5\% percentiles range of the 2000 MCMC samples. The dotted vertical line represents the time of prediction $t$.} \label{fig:UPDRS}
\end{figure}

Parameter estimates based on Model 1 are presented in Table~\ref{tab:inf} and Web Table \ref{tab:S2} (outcome-specific parameters only). To illustrate the subject-specific predictions, we set aside two patients from the DATATOP study and predict their longitudinal trajectories as well as the probability of functional disability at a clinically relevant future time point, conditional on their available measurements. A more severe Patient 169 with clinically worse longitudinal measures and earlier development of functional disability as well as a less severe Patient 718 are selected. Patient 169 had 8 visits with mean UPDRS 42.6 (SD 7.7), median HY 2, median SEADL 80, and developed functional disability at month 16. In contrast, Patient 718 had 9 visits with mean UPDRS 15.6 (SD 3.1), median HY 1, median SEADL 95, and was censored at month 21. Figure~\ref{fig:UPDRS} displays the predicted UPDRS trajectories for these two patients, based on different amounts of data. When only baseline measurements are used for prediction, the predicted UPDRS trajectory is biased with wide uncertainty band. For example, Patient 169 had a relatively low baseline UPDRS value of 33 and our model based only on baseline measurements tends to underpredict the future UPDRS trajectory ($t_i=0$, the first plot in upper panels). However, Patient 169's higher UPDRS values of 41 and 40 at months 1 and 3, respectively, subsequently shift up the prediction and tend to overpredict the future trajectory ($t_i=3$ months, the second plot in upper panels). By using more follow-up data, predictions are closer to the true observed values and the 95\% uncertainty band is narrower ($t_i=6$ or $12$ months, the last two plots in upper panels). Patient 169's predicted UPDRS values after 12 months are above 40 and increase rapidly, indicating a higher risk of functional disability in the near future. In comparison, the predicted UPDRS values for Patient 718 are relatively stable because his/her observed UPDRS values are relatively stable.

\begin{table}[h!tbp]
\caption{Parameter estimates for the DATATOP study from Model 1.} \label{tab:inf}
\centering
\begin{tabular}{lrrrr}
  \hline
    & Mean & SD & \multicolumn{2}{c}{95\% CI} \\
  \hline
  \multicolumn{5}{l}{For latent disease severity} \\
  Int & $-$0.738 & 0.338 & $-$1.385 & $-$0.081 \\
  Duration (months) & 0.021 & 0.004 & 0.014 & 0.028 \\
  Age (years) & 0.024 & 0.005 & 0.014 & 0.035 \\
  Trt (deprenyl) & $-$0.108 & 0.099 & $-$0.304 & 0.099 \\
  Time (months) & 0.021 & 0.025 & $-$0.028 & 0.070 \\
  Trt $\times$ Time & $-$0.089 & 0.010 & $-$0.109 & $-$0.071 \\
  $\rho$ & 0.310 & 0.044 & 0.226 & 0.393 \\
  $\sigma_1$ & 1.328 & 0.051 & 1.230 & 1.430 \\
  $\sigma_2$ & 0.116 & 0.006 & 0.104 & 0.128 \\
  $\sigma_{\varepsilon}$ & 5.081 & 0.074 & 4.933 & 5.226 \\
  \\
  \multicolumn{5}{l}{For survival process} \\
  Duration (months) & $-$0.009 & 0.004 & $-$0.017 & $-$0.002 \\
  Age (years) & $-$0.034 & 0.006 & $-$0.045 & $-$0.024 \\
  Trt (deprenyl) & $-$0.608 & 0.118 & $-$0.846 & $-$0.375 \\
  $\nu$ & 0.692 & 0.039 & 0.618 & 0.769 \\
   \hline
\end{tabular}
\end{table}

\begin{figure}[h!tbp]
\centering
\includegraphics[width=1\textwidth,angle=0]{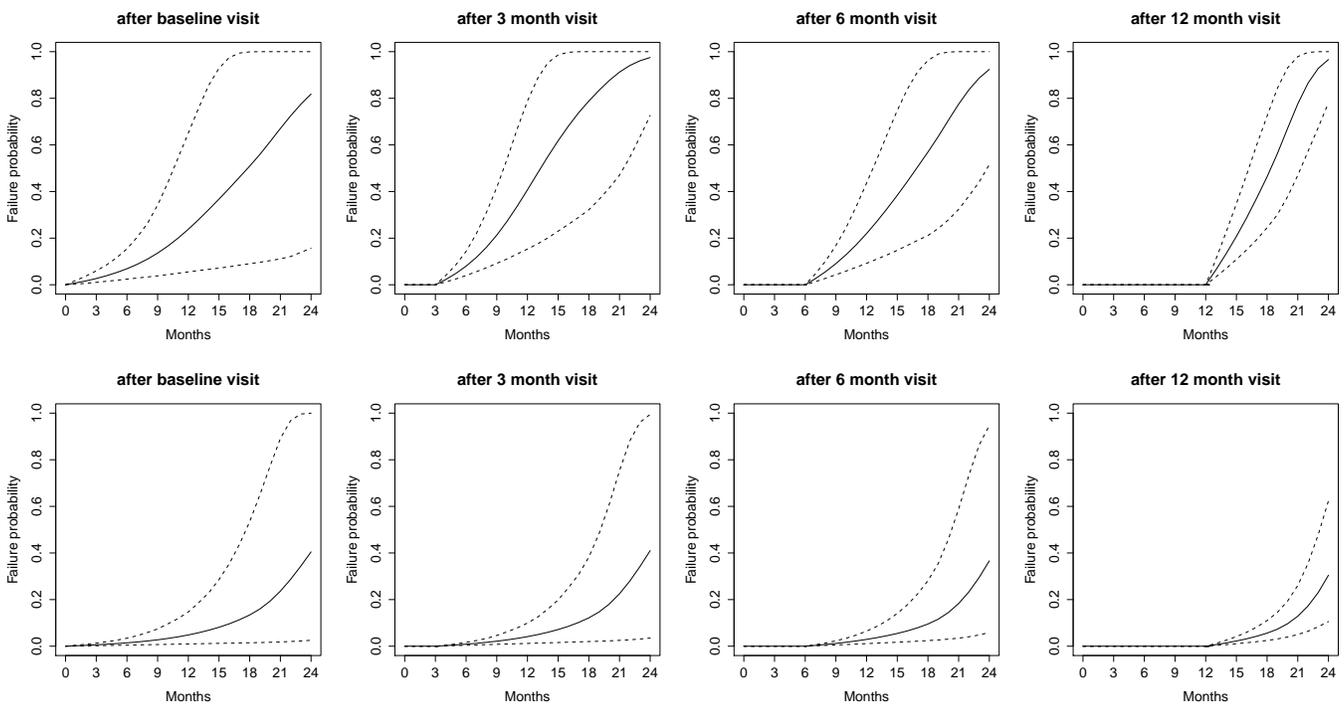}
\caption{Predicted conditional failure probability for Patient 169 (upper panels) and Patient 718 (lower panels). Solid line is the mean of 2000 MCMC samples. Dashed lines are the 2.5\% and 97.5\% percentiles range of the MCMC samples.} \label{fig:failure}
\end{figure}

The predicted probability being in each category for outcomes HY and SEADL are presented in Web Figures \ref{fig:S1} and \ref{fig:S2}, respectively. Please refer to the \ref{sec:WebSupp} for the interpretation. Besides the predictions of longitudinal trajectories, it is more of clinical interest for patients and clinicians to know the probability of functional disability before time $t'>t$: $\pi_i(t'|t)$, conditional on the patient's longitudinal profiles up to time $t$ and the fact that he/she did not have functional disability up to time $t$. The predicted probabilities for Patients 169 and 718 based on various amount of data are presented in Figure~\ref{fig:failure}. A similar pattern is that the prediction becomes more accurate if more data are used. With such predictions, clinicians are able to precisely track the health condition of each patient and make better informed decisions individually. For example, based on the first 12 months' data, for Patient 169, the predicted probabilities in the next 3, 6, 9 and 12 months are 0.21, 0.46, 0.78 and 0.97 (the last plot of upper panels), while for Patient 718, the probabilities are 0.02, 0.06, 0.13 and 0.30 (the last plot of lower panels). Patient 169 has higher risk of functional disability in the next few months and clinicians may consider more invasive treatments to control the disease symptoms before the functional disability is developed.

\begin{figure}[h!tbp]
\centering
\includegraphics[width=1\textwidth,angle=0]{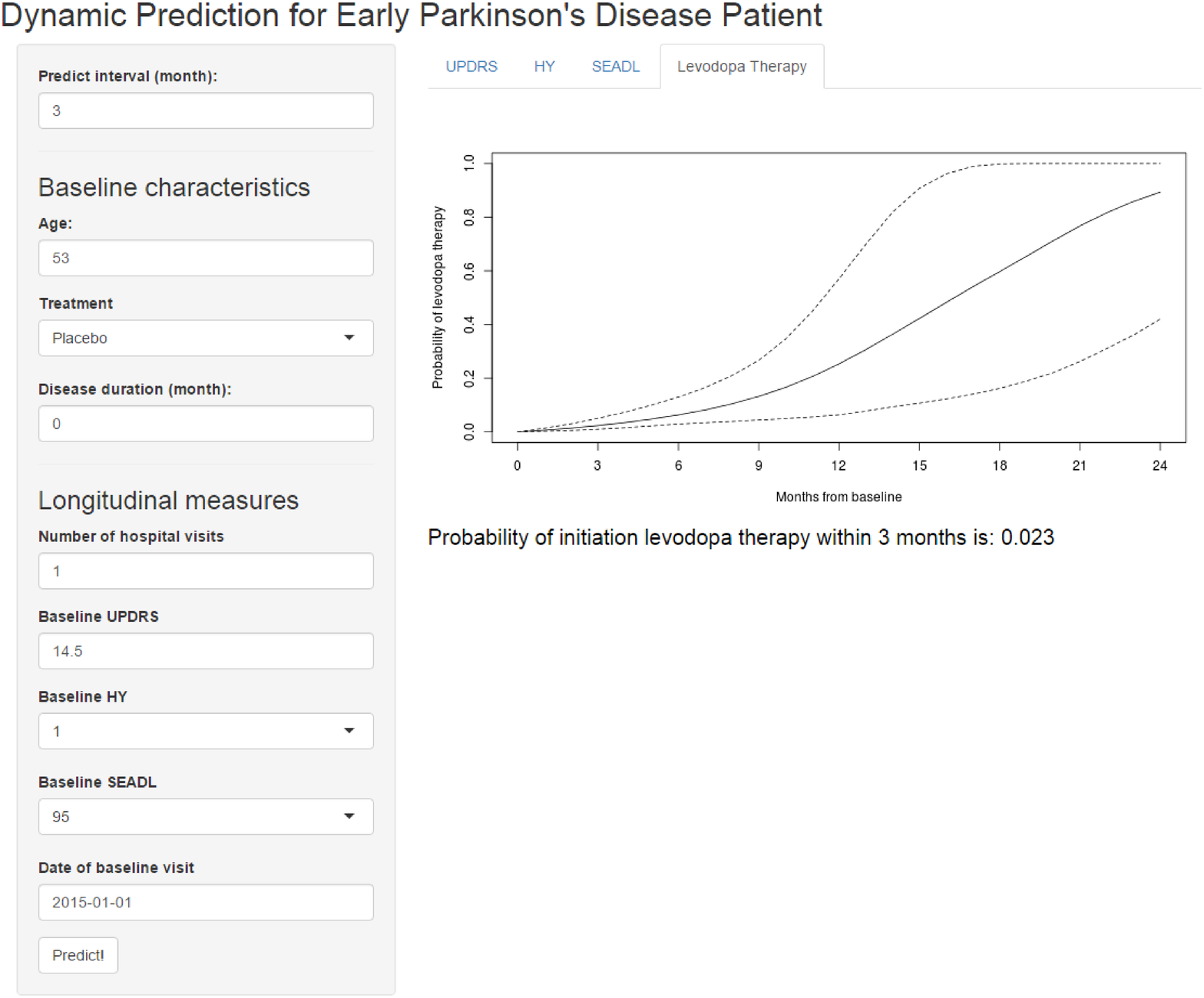}
\caption{A screenshot of the web-based calculator for prediction.} \label{fig:S3}
\end{figure}

To facilitate the personalized dynamic predictions in clinical setting, we develop a web-based calculator available at \url{https://kingjue.shinyapps.io/dynPred_PD}. A screenshot of the user interface is presented in Web Figure \ref{fig:S3}. The calculator requires as input the PD patients' baseline characteristics and their longitudinal outcome values up to the present time. The online calculator will then produce time-dependent predictions of future health outcomes trajectories and the probability of functional disability, in addition to the $95\%$ uncertainty bands. Moreover, additional data generated from more follow-up visits can be input to obtain updated predictions. The calculator is a user friendly and easily accessible tool to provide clinicians with dynamically-updated patient-specific future health outcome trajectories, risk predictions, and the associated uncertainty. Such a translational tool would be relevant both for clinicians to make informed decisions on therapy selection and for patients to better manage risks.

\section{Simulation studies}\label{sec:simulation}
In this section, we conduct an extensive simulation study to investigate the prediction performance of the probability $\pi(t'|t)$ using the proposed Model 1. We generate 200 datasets with samples size $n = 800$ subjects and six visits, i.e., baseline and five follow-up visits ($J_i=6$), with the time vector $\boldsymbol{t}_i = (t_{i1}, t_{i2}, \ldots, t_{i6})' = (0, 3, 6, 12, 18, 24)$. The simulated data structure is similar to the motivating DATATOP study, and it includes one continuous outcome and two ordinal outcomes (each with 7 categories).

Data are generated from the following models: $\theta_i(t_{ij}) = \beta_0 + \beta_1x_{i1} + \beta_2t_{ij} + \beta_3x_{i1}t_{ij} + u_{i0} + u_{i1}t_{ij}$ and $h_i(t) = h_0\exp\{\gamma x_{i2} + \nu \theta_i(t)\}$, where the longitudinal and survival submodels share the latent variable as in proposed Model 1. Covariate $x_{i1}$ takes value 0 or 1 each with probability 0.5 to mimic treatment assignment and covariate $x_{i2}$ is randomly sampled integer from 30 to 80 to mimic age. We set coefficients $\boldsymbol{\beta} = (\beta_0, \beta_1, \beta_2, \beta_3)' = (-1, -0.2, 0.8, -0.2)'$, $\gamma=-0.12$ and $\nu=0.75$. For simplicity, baseline hazard is assumed to be constant with $h_0 = 0.1$. Parameters for the continuous outcome are $a_1 = 15$, $b_1=7$ and $\sigma_\varepsilon = 5$. Parameters for the ordinal outcomes are $\boldsymbol{a}_2 = (0, 1, 2, 4, 5, 6)$, $\boldsymbol{a}_3 = (-1, 1, 3, 4, 6, 8)$, $b_2=1$ and $b_3=1.2$. We assume that random effects vector $\boldsymbol{u}_i = (u_{i0}, u_{i1})'$ follows a multivariate normal distribution $N_2(0, \boldsymbol{\Sigma})$, where $\boldsymbol{\Sigma} = \{(\sigma_1^2, \rho\sigma_1\sigma_2), (\rho\sigma_1\sigma_2, \sigma_2^2) \}$ with $\sigma_1=1.5$, $\sigma_2=0.15$ and $\rho=0.4$. The independent censoring time is sampled from $\textnormal{Uniform}(10, 24)$.

From each simulated dataset, we randomly select 600 subjects as the training dataset and set aside the remaining 200 subjects as the validation dataset. Web Table \ref{tab:S3} displays bias (the average of the posterior means minus the true values), standard deviation (SD, the standard deviation of the posterior means), coverage probabilities (CP) of 95\% equal tail credible intervals (CI), and root mean squared error (RMSE) of model inference based on the training dataset. The results suggest that the model fitting based on the training dataset provides parameter estimates with very small biases and RMSE and the CP being close to the nominal level 0.95. Using MCMC samples from the fitted model and available measurements up to time $t$, we make prediction of $\pi_i(t'|t)$ for each subject in the validation dataset.

Web Table \ref{tab:S4} compares the time-dependent AUC based on various amount of data from Model 1, Model JM and naive Cox model. When 3 or 6 months data are available, Model 1 outperforms Model JM and Cox with high discriminating capability and higher AUC values above 0.9. In general, AUC is increasing with more available data, e.g., $\textnormal{AUC}(3, 12) = 0.920$ and $\textnormal{AUC}(6, 12) = 0.930$.

\begin{figure}[h!tbp]
\centering
\includegraphics[width=1\textwidth,angle=0]{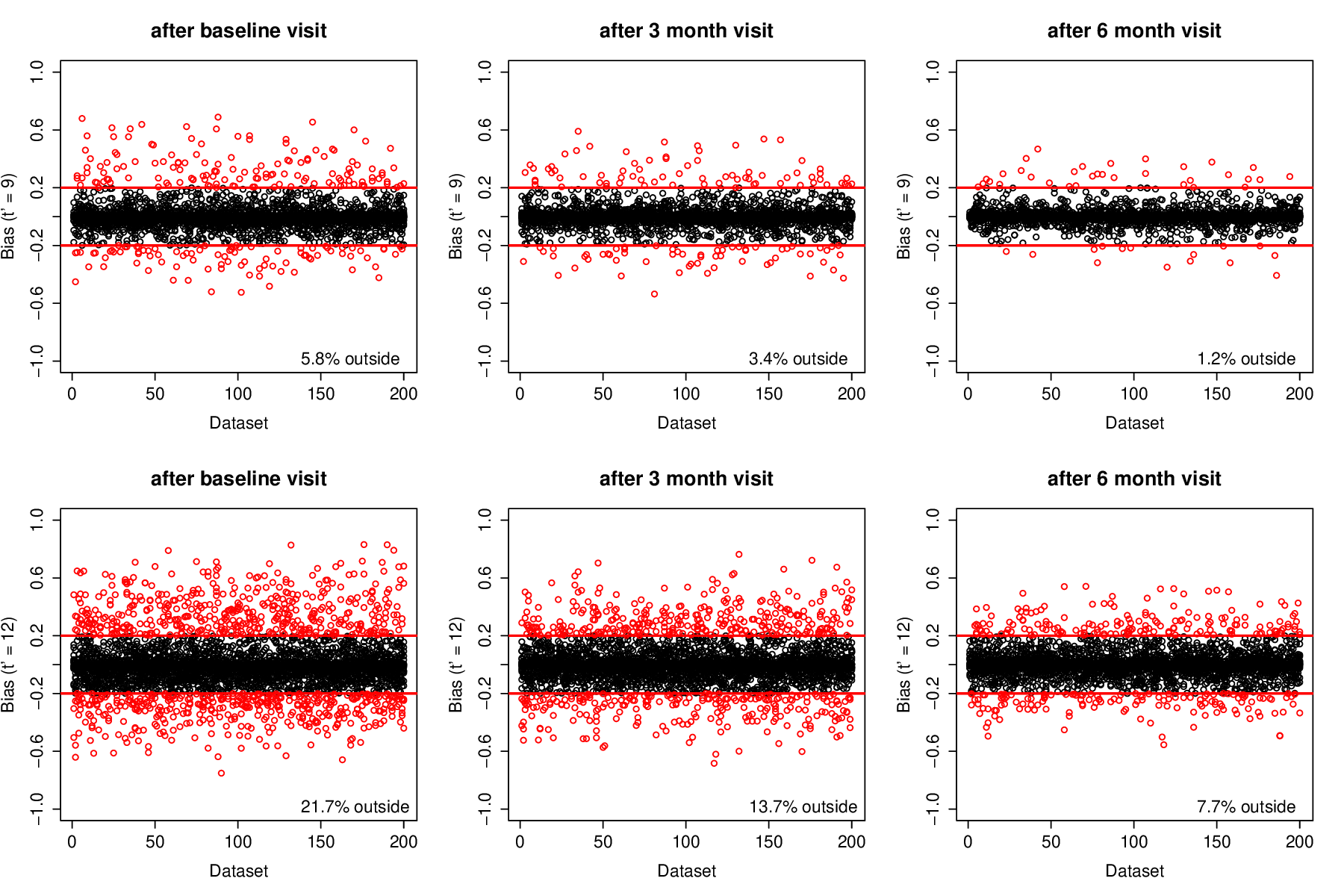}
\caption{Bias between the predicted failure probability $\widehat{\pi}_i(t'|\boldsymbol{y}_i^{\{t\}}, \boldsymbol{X}_i^{\{t\}})$ with true failure probability when $t'=9$ (upper panels) and $t'=12$ (lower panels) for 20 randomly selected subjects from each of the 200 simulation datasets.} \label{fig:S4}
\end{figure}

From each of the 200 simulation datasets, we randomly select 20 subjects to plot the bias between the predicted event probability $\pi(t'|t)$ from Model 1 and the true event probability with $t'=9$ (upper panels) and $t'=12$ (lower panels) in Web Figure \ref{fig:S4}. When more data are available, bias is decreasing as more bias is within the region of $[-0.2, 0.2]$. For example, with only baseline data, 5.8\% and 21.7\% of bias for the predictions of $\pi(t'=9|t=0)$ and $\pi(t'=12|t=0)$, respectively, are outside the range. With up to three months' data, 3.4\% and 13.7\% of bias for the predictions of $\pi(t'=9|t=3)$ and $\pi(t'=12|t=3)$, respectively, are outside the range. With up to six months' data, the prediction is precise with only 1.2\% and 7.7\% of bias for the prediction of $\pi(t'=9|t=6)$ and $\pi(t'=12|t=6)$, respectively, being outside the range.

\section{Discussion} \label{sec:discussion}
Multiple longitudinal outcomes are often collected in clinical trials of complex diseases such as Parkinson's disease (PD) to better measure different aspects of disease impairment. However, both theoretical and computational complexity in modeling multiple longitudinal outcomes often restrict researchers to a univariate longitudinal outcome. Without careful analysis of the entire data, pace of treatment discovery can be dramatically slowed down. 

In this article, we first propose a joint model that consists of a semiparametric multilevel latent trait model (MLLTM) for the multiple longitudinal outcomes by introducing a continuous latent variable to represent patients' underlying disease severity, and a survival submodel for the event time data. The latent variable modeling effectively reduces the number of outcomes and has improved computational feasibility and model interpretability. Next we develop the process of making personalized dynamic predictions of future outcome trajectories and risks of target event. Extensive simulation studies suggest that the predictions are accurate with high AUC and small bias. We apply the method to the motivating DATATOP study. The proposed joint models can efficiently utilize the multivariate longitudinal outcomes of mixed types, as well as the survival process to make correct predictions for new subjects. When new measurements are available, predictions can be dynamically updated and become more accurate and efficient. A web-based calculator is developed as a supplemental tool for PD clinicians to monitor their patients' disease progression. For subjects with high predicted risk of functional disability in the near future, clinicians may consider more targeted treatment to defer the initiation of levodopa therapy because of its association with motor complications and notable adverse events \citep{Brooks2008NDT}. Although the dynamic prediction framework has utilized only three longitudinal outcomes in the DATATOP study, it can be broadly applied to similar studies with more longitudinal outcomes.

There are some limitations in our proposed dynamic prediction framework that we will address in the future study. First, the semiparametric MLLTM submodel assumes a univariate latent variable (unidimensional assumption), which may be reasonable for small number of outcomes. However, for large number of longitudinal outcomes, multiple latent variables may be required to fully represent the true disease severity across different domains impaired by PD. We will develop a multidimensional latent trait model that allows multiple latent variables. Second, \cite{Proust-Lima2013BJMSP} and \cite{Proust2016SIM} proposed a flexible multivariate longitudinal model that can handle mixed outcomes, including bounded and non-Gaussian continuous outcomes. In contrast, our model~\eqref{eqn:MLLTM_continuous} only applied to normally distributed continuous outcomes. In our future research, we would like to extend the dynamic prediction framework to accommodate more general continuous outcomes including bounded and non-Gaussian variables. Third, we have chosen multivariate normal distribution for the random effects vector because it is flexible in modeling the covariance structure within and between longitudinal measures of patients and it has meaningful interpretation on correlation. In fact, misspecification of random effects and residuals has little impact on the parameters that are not associated with the random effects \citep{Jacqmin2007CSDA, Rizopoulos2008Biometrika, Mcculloch2011SS}. The impact of misspecification in the proposed modeling framework warrants further investigation. Alternatively, we will relax the normality assumption by considering Bayesian non-parametric (BNP) framework based on Dirichlet process mixture \citep{Escobar1994JASA}.

Equation~\eqref{eqn:MLLTM_ordi} for ordinal outcome requires the proportional odds assumption. Statistical tests to evaluate this assumption in the traditional ordinal logistic regression have been criticized for having a tendency to reject the null hypothesis, when the assumption holds \citep{Harrell2015Book}. Tests of the proportional odds assumption in the longitudinal latent variable setting are not well established, and the consequence of violating the assumption is unclear and is worth future examination. Three different functional forms of joint models that allow various association between the longitudinal and event time responses are examined and they provide comparable predictions in the DATATOP study. Instead of selecting a final model in terms of simplicity and easy interpretation, a Bayesian model averaging (BMA) approach to combine joint models with different association structures \citep{Rizopoulos2014JASA} will be investigated in future study. In addition, missed visits and missing covariates exist in the DATATOP study. In this article, we assume that they are missing at random (MAR). However, the missing data issue becomes more complicated in prediction model framework because it can impact both the model inference (missing data in the training dataset) and dynamic prediction process (e.g., the new subject only has measurements of UPDRS and HY, but not SEADL). How to address this issue in the proposed prediction framework is an important direction of future research. Moreover, the online calculator is based on the DATATOP study, which may not represent PD patients at all stages and from all populations. Nonetheless, the large and carefully studied group of patients provide an important resource to study the clinical expression of PD. We will continue to improve the calculator by including more heterogeneous PD patients from different studies.

\section*{Acknowledgements}
Sheng Luo's research was supported by the National Institute of Neurological Disorders and Stroke under Award Numbers R01NS091307 and 5U01NS043127. The authors acknowledge the Texas Advanced Computing Center (TACC) for providing high-performing computing resources.

\bibliographystyle{authordate1}
\bibliography{literatures_DP}

\beginsupplement
\newpage
\section*{Web Supplement} \label{sec:WebSupp}
\begin{table}[h!tbp]
\centering
\footnotesize
\caption{Area under the ROC curve and Brier score (BS) for the DATATOP study.} \label{tab:S1}
\begin{tabular}{rrcccccccccc}
  \hline
     & & \multicolumn{2}{c}{Model 1} & \multicolumn{2}{c}{Model 2} & \multicolumn{2}{c}{Model 3} & \multicolumn{2}{c}{Model JM} & \multicolumn{2}{c}{Cox} \\
  \hline
   $t$ & $t'$ & \multicolumn{10}{c}{AUC (95\% CI)} \\
  \hline
  3 & 9 & \multicolumn{2}{c}{0.754} & \multicolumn{2}{c}{0.759} & \multicolumn{2}{c}{0.761} & \multicolumn{2}{c}{0.757} & \multicolumn{2}{c}{0.736} \\
    &   & \multicolumn{2}{c}{(0.703, 0.802)} & \multicolumn{2}{c}{(0.710, 0.805)} & \multicolumn{2}{c}{(0.713, 0.806)} & \multicolumn{2}{c}{(0.708, 0.802)} & \multicolumn{2}{c}{(0.681, 0.786)} \\
  & 12 & \multicolumn{2}{c}{0.744} & \multicolumn{2}{c}{0.744} & \multicolumn{2}{c}{0.744} & \multicolumn{2}{c}{0.739} & \multicolumn{2}{c}{0.725} \\
    &   & \multicolumn{2}{c}{(0.702, 0.785)} & \multicolumn{2}{c}{(0.702, 0.784)} & \multicolumn{2}{c}{(0.701, 0.785)} & \multicolumn{2}{c}{(0.696, 0.780)} & \multicolumn{2}{c}{(0.682, 0.768)} \\
  & 15 & \multicolumn{2}{c}{0.744} & \multicolumn{2}{c}{0.742} & \multicolumn{2}{c}{0.744} & \multicolumn{2}{c}{0.726} & \multicolumn{2}{c}{0.719} \\
    &   & \multicolumn{2}{c}{(0.702, 0.783)} & \multicolumn{2}{c}{(0.702, 0.780)} & \multicolumn{2}{c}{(0.704, 0.784)} & \multicolumn{2}{c}{(0.682, 0.768)} & \multicolumn{2}{c}{(0.677, 0.760)} \\
  & 18 & \multicolumn{2}{c}{0.775} & \multicolumn{2}{c}{0.766} & \multicolumn{2}{c}{0.772} & \multicolumn{2}{c}{0.728} & \multicolumn{2}{c}{0.720} \\
    &   & \multicolumn{2}{c}{(0.731, 0.819)} & \multicolumn{2}{c}{(0.719, 0.811)} & \multicolumn{2}{c}{(0.723, 0.814)} & \multicolumn{2}{c}{(0.679, 0.772)} & \multicolumn{2}{c}{(0.673, 0.765)} \\
	
  6 & 9 & \multicolumn{2}{c}{0.789} & \multicolumn{2}{c}{0.806} & \multicolumn{2}{c}{0.806} & \multicolumn{2}{c}{0.770} & \multicolumn{2}{c}{0.721} \\
    &   & \multicolumn{2}{c}{(0.717, 0.851)} & \multicolumn{2}{c}{(0.739, 0.865)} & \multicolumn{2}{c}{(0.740, 0.864)} & \multicolumn{2}{c}{(0.699, 0.834)} & \multicolumn{2}{c}{(0.642, 0.795)} \\
  & 12 & \multicolumn{2}{c}{0.764} & \multicolumn{2}{c}{0.778} & \multicolumn{2}{c}{0.775} & \multicolumn{2}{c}{0.732} & \multicolumn{2}{c}{0.705} \\
    &   & \multicolumn{2}{c}{(0.717, 0.809)} & \multicolumn{2}{c}{(0.731, 0.821)} & \multicolumn{2}{c}{(0.729, 0.820)} & \multicolumn{2}{c}{(0.679, 0.782)} & \multicolumn{2}{c}{(0.651, 0.757)} \\
  & 15 & \multicolumn{2}{c}{0.763} & \multicolumn{2}{c}{0.771} & \multicolumn{2}{c}{0.771} & \multicolumn{2}{c}{0.725} & \multicolumn{2}{c}{0.697} \\
    &   & \multicolumn{2}{c}{(0.716, 0.807)} & \multicolumn{2}{c}{(0.726, 0.814)} & \multicolumn{2}{c}{(0.726, 0.814)} & \multicolumn{2}{c}{(0.678, 0.771)} & \multicolumn{2}{c}{(0.648, 0.743)} \\
  & 18 & \multicolumn{2}{c}{0.786} & \multicolumn{2}{c}{0.773} & \multicolumn{2}{c}{0.769} & \multicolumn{2}{c}{0.726} & \multicolumn{2}{c}{0.701} \\
    &   & \multicolumn{2}{c}{(0.736, 0.832)} & \multicolumn{2}{c}{(0.726, 0.821)} & \multicolumn{2}{c}{(0.720, 0.815)} & \multicolumn{2}{c}{(0.674, 0.776)} & \multicolumn{2}{c}{(0.650, 0.751)} \\
	
  12 & 15 & \multicolumn{2}{c}{0.766} & \multicolumn{2}{c}{0.787} & \multicolumn{2}{c}{0.782} & \multicolumn{2}{c}{0.695} & \multicolumn{2}{c}{0.647} \\
    &   & \multicolumn{2}{c}{(0.701, 0.828)} & \multicolumn{2}{c}{(0.716, 0.850)} & \multicolumn{2}{c}{(0.710, 0.849)} & \multicolumn{2}{c}{(0.623, 0.768)} & \multicolumn{2}{c}{(0.563, 0.727)} \\
  & 18 & \multicolumn{2}{c}{0.758} & \multicolumn{2}{c}{0.739} & \multicolumn{2}{c}{0.723} & \multicolumn{2}{c}{0.700} & \multicolumn{2}{c}{0.663} \\
    &   & \multicolumn{2}{c}{(0.684, 0.824)} & \multicolumn{2}{c}{(0.671, 0.808)} & \multicolumn{2}{c}{(0.651, 0.790)} & \multicolumn{2}{c}{(0.631, 0.774)} & \multicolumn{2}{c}{(0.592, 0.732)} \\
   \hline
   $t$ & $t'$ & \multicolumn{10}{c}{BS (95\% CI)} \\
   \hline
  3 & 9 & \multicolumn{2}{c}{0.136} & \multicolumn{2}{c}{0.138} & \multicolumn{2}{c}{0.139} & \multicolumn{2}{c}{0.140} & \multicolumn{2}{c}{0.139} \\
    &   & \multicolumn{2}{c}{(0.116, 0.155)} & \multicolumn{2}{c}{(0.119, 0.158)} & \multicolumn{2}{c}{(0.119, 0.158)} & \multicolumn{2}{c}{(0.120, 0.159)} & \multicolumn{2}{c}{(0.122, 0.156)} \\
  & 12 & \multicolumn{2}{c}{0.204} & \multicolumn{2}{c}{0.200} & \multicolumn{2}{c}{0.200} & \multicolumn{2}{c}{0.203} & \multicolumn{2}{c}{0.203} \\
    &   & \multicolumn{2}{c}{(0.181, 0.226)} & \multicolumn{2}{c}{(0.180, 0.221)} & \multicolumn{2}{c}{(0.180, 0.220)} & \multicolumn{2}{c}{(0.182, 0.225)} & \multicolumn{2}{c}{(0.184, 0.221)} \\
  & 15 & \multicolumn{2}{c}{0.216} & \multicolumn{2}{c}{0.212} & \multicolumn{2}{c}{0.211} & \multicolumn{2}{c}{0.218} & \multicolumn{2}{c}{0.212} \\
    &   & \multicolumn{2}{c}{(0.192, 0.240)} & \multicolumn{2}{c}{(0.191, 0.234)} & \multicolumn{2}{c}{(0.191, 0.231)} & \multicolumn{2}{c}{(0.195, 0.242)} & \multicolumn{2}{c}{(0.193, 0.232)} \\
  & 18 & \multicolumn{2}{c}{0.171} & \multicolumn{2}{c}{0.163} & \multicolumn{2}{c}{0.167} & \multicolumn{2}{c}{0.186} & \multicolumn{2}{c}{0.185} \\
    &   & \multicolumn{2}{c}{(0.148, 0.196)} & \multicolumn{2}{c}{(0.142, 0.186)} & \multicolumn{2}{c}{(0.147, 0.188)} & \multicolumn{2}{c}{(0.162, 0.211)} & \multicolumn{2}{c}{(0.164, 0.207)} \\
	
  6 & 9 & \multicolumn{2}{c}{0.078} & \multicolumn{2}{c}{0.078} & \multicolumn{2}{c}{0.078} & \multicolumn{2}{c}{0.081} & \multicolumn{2}{c}{0.094} \\
    &   & \multicolumn{2}{c}{(0.062, 0.095)} & \multicolumn{2}{c}{(0.062, 0.095)} & \multicolumn{2}{c}{(0.062, 0.095)} & \multicolumn{2}{c}{(0.064, 0.099)} & \multicolumn{2}{c}{(0.080, 0.108)} \\
  & 12 & \multicolumn{2}{c}{0.159} & \multicolumn{2}{c}{0.154} & \multicolumn{2}{c}{0.154} & \multicolumn{2}{c}{0.164} & \multicolumn{2}{c}{0.173} \\
    &   & \multicolumn{2}{c}{(0.137, 0.180)} & \multicolumn{2}{c}{(0.134, 0.174)} & \multicolumn{2}{c}{(0.134, 0.174)} & \multicolumn{2}{c}{(0.143, 0.186)} & \multicolumn{2}{c}{(0.155, 0.191)} \\
  & 15 & \multicolumn{2}{c}{0.183} & \multicolumn{2}{c}{0.178} & \multicolumn{2}{c}{0.178} & \multicolumn{2}{c}{0.194} & \multicolumn{2}{c}{0.194} \\
    &   & \multicolumn{2}{c}{(0.160, 0.207)} & \multicolumn{2}{c}{(0.157, 0.199)} & \multicolumn{2}{c}{(0.158, 0.199)} & \multicolumn{2}{c}{(0.171, 0.218)} & \multicolumn{2}{c}{(0.176, 0.214)} \\
  & 18 & \multicolumn{2}{c}{0.158} & \multicolumn{2}{c}{0.154} & \multicolumn{2}{c}{0.159} & \multicolumn{2}{c}{0.175} & \multicolumn{2}{c}{0.175} \\
    &   & \multicolumn{2}{c}{(0.133, 0.183)} & \multicolumn{2}{c}{(0.132, 0.178)} & \multicolumn{2}{c}{(0.138, 0.183)} & \multicolumn{2}{c}{(0.151, 0.201)} & \multicolumn{2}{c}{(0.155, 0.197)} \\
	
  12 & 15 & \multicolumn{2}{c}{0.108} & \multicolumn{2}{c}{0.103} & \multicolumn{2}{c}{0.102} & \multicolumn{2}{c}{0.124} & \multicolumn{2}{c}{0.155} \\
    &   & \multicolumn{2}{c}{(0.084, 0.133)} & \multicolumn{2}{c}{(0.082, 0.125)} & \multicolumn{2}{c}{(0.080, 0.125)} & \multicolumn{2}{c}{(0.100, 0.150)} & \multicolumn{2}{c}{(0.136, 0.174)} \\
  & 18 & \multicolumn{2}{c}{0.149} & \multicolumn{2}{c}{0.147} & \multicolumn{2}{c}{0.153} & \multicolumn{2}{c}{0.161} & \multicolumn{2}{c}{0.163} \\
    &   & \multicolumn{2}{c}{(0.121, 0.178)} & \multicolumn{2}{c}{(0.121, 0.174)} & \multicolumn{2}{c}{(0.126, 0.179)} & \multicolumn{2}{c}{(0.132, 0.192)} & \multicolumn{2}{c}{(0.139, 0.187)} \\
   \hline
\end{tabular}
\end{table}

\begin{table}[h!tbp]
\caption{The outcome-specific parameter estimates for the DATATOP study from Model 1.} \label{tab:S2}
\centering
\begin{tabular}{lrrrr}
  \hline
    & Mean & SD & \multicolumn{2}{c}{95\% CI} \\
  \hline
  \multicolumn{5}{l}{For UPDRS} \\
  $a_1$ & 17.247 & 0.341 & 16.563 & 17.902 \\
  $b_1$ & 7.624 & 0.207 & 7.251 & 8.035 \\
  \\
  \multicolumn{5}{l}{For HY} \\
  $a_{22}$ & 0.995 & 0.036 & 0.927 & 1.066 \\
  $a_{23}$ & 4.087 & 0.079 & 3.935 & 4.243 \\
  $a_{24}$ & 6.340 & 0.129 & 6.087 & 6.593 \\
  \\
  \multicolumn{5}{l}{For SEADL} \\
  $a_{31}$ & $-$1.462 & 0.076 & $-$1.610 & $-$1.311 \\
  $a_{32}$ & 0.583 & 0.071 & 0.450 & 0.720 \\
  $a_{33}$ & 3.008 & 0.088 & 2.838 & 3.181 \\
  $a_{34}$ & 3.860 & 0.096 & 3.679 & 4.051 \\
  $a_{35}$ & 6.020 & 0.132 & 5.770 & 6.283 \\
  $a_{36}$ & 6.851 & 0.151 & 6.558 & 7.150 \\
  $a_{37}$ & 8.474 & 0.203 & 8.082 & 8.874 \\
  $b_3$ & 1.270 & 0.045 & 1.187 & 1.363 \\
   \hline
\end{tabular}
\end{table}

\section*{Predicted Probability for Ordinal Outcomes}

\begin{figure}[h!tbp]
\centering
\includegraphics[width=1\textwidth,angle=0]{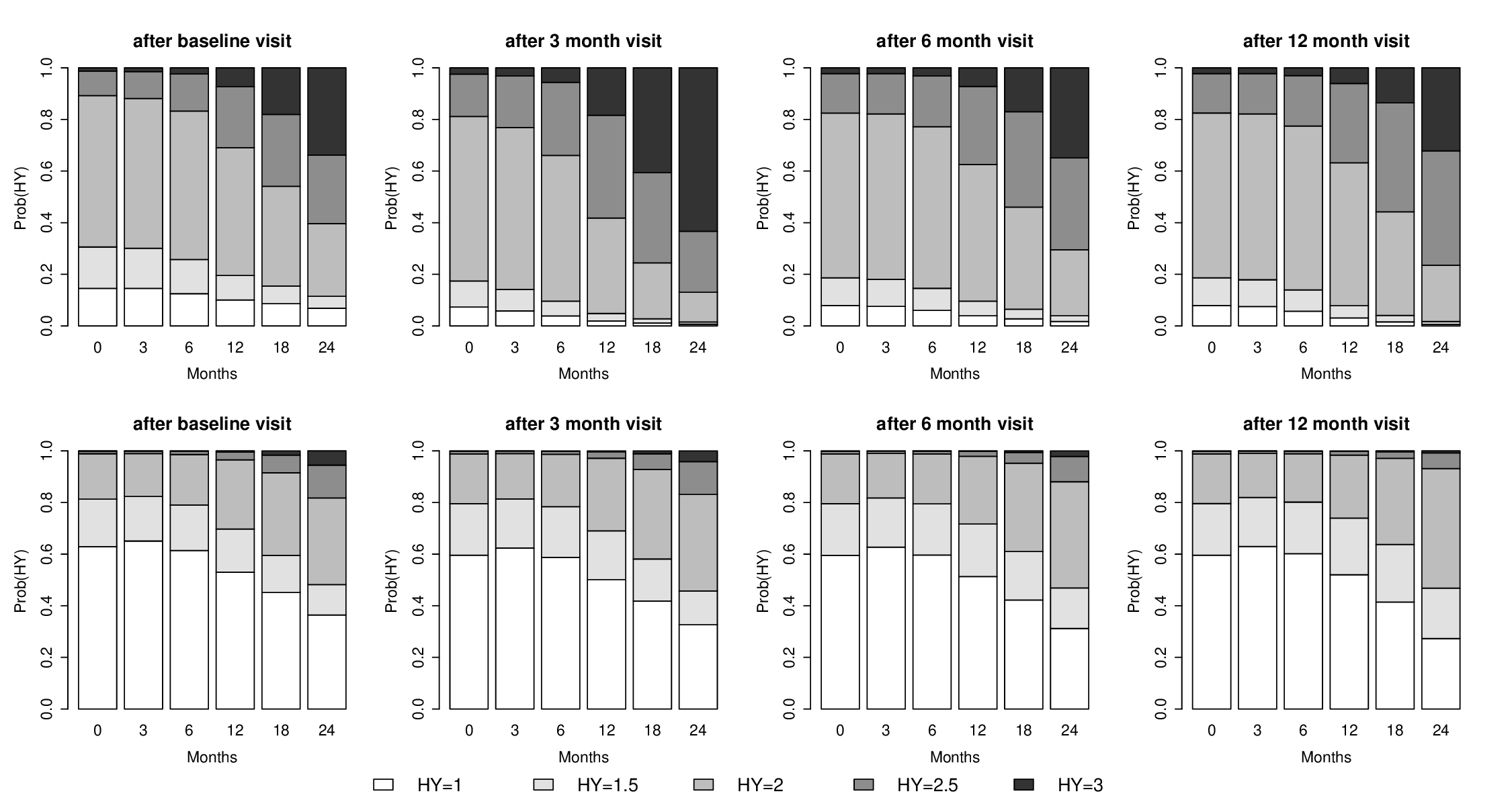}
\caption{Predicted probability of being in each HY category for Patient 169 (upper panels) and Patient 718 (lower panels). Patient 169 had HY measurements equal to $2$ at all 8 visits at months 0, 1, 3, 6, 9, 12, 15, and 16, while Patient 718 had HY measurements equal to $1$ at all 9 visits at months 0, 1, 3, 6, 9, 12, 15, and 18.
  } \label{fig:S1}
\end{figure}

\begin{figure}[h!tbp]
\centering
\includegraphics[width=1\textwidth,angle=0]{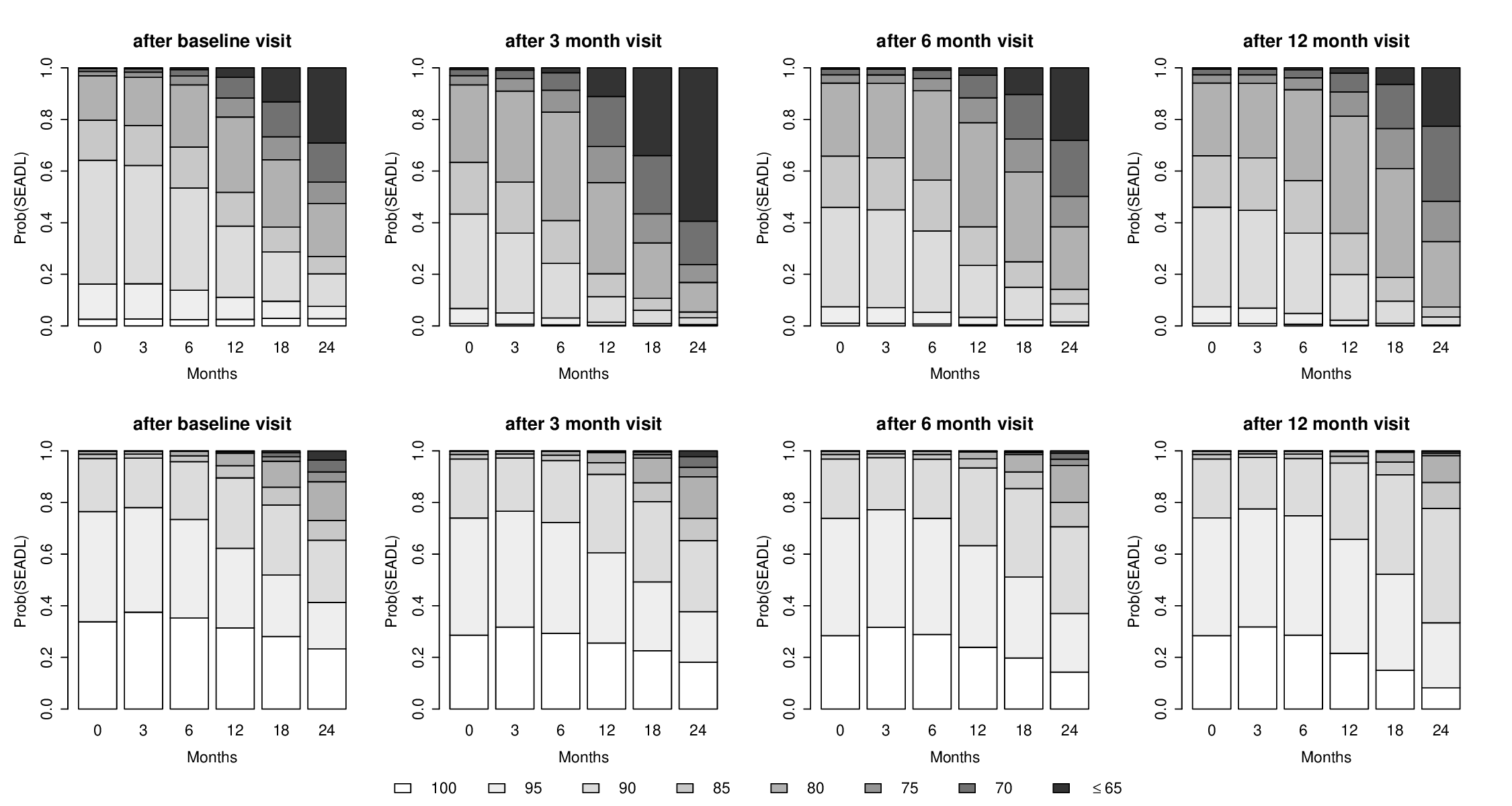}
\caption{Predicted probability of SEADL to be observed in a given category for Subject 169 (upper panels) and Subject 718 (lower panels). Observed categories of SEADL for Subject 169 in the 8 follow-up visits are 90, 80, 80, 90, 80, 80, 80, 80 and for Subject 718 in the 9 visits are 95, 95, 95, 95, 90, 95, 95, 95, 95.
  } \label{fig:SEADL} \label{fig:S2}
\end{figure}

The predicted probability being in each category for outcome HY is presented in Figure~\ref{fig:S1}. For example, Patient 169 had HY measurements equal to $2$ at all visits. When only the baseline data are used for prediction (the first plot in upper panels), our model tends to underpredict the disease progression by assigning sizable probabilities to the less severe HY categories 1 and 1.5 even at the end of the study, possibly due to low baseline UPDRS value of 33. After month 3 visit (the second plot in upper panels), our model overpredicts disease progression by assigning abnormally high probability to the severe category 3, possibly due to higher UPDRS values at months 1 and 3. However, using the first 6 or 12 months' data (the last two plots in upper panels), our model has good fit by correctly assigning the largest posterior probability to HY category 2 for all visits from baseline to month 12. Moreover, our model properly assigns higher probabilities to more severe categories 2.5 and 3 and negligible probabilities to less severe categories 1 and 1.5 for visits after month 12, due to the deteriorating UPDRS measure. Similar interpretation can be made to the predicted probability of being in each SEADL category displayed in Figure~\ref{fig:S2}.

\begin{table}[h!tbp]
\caption{Simulation results using the training dataset.} \label{tab:S3}
\centering
\begin{tabular}{lrrrr}
  \hline
 & BIAS & SD & CP & RMSE \\
  \hline
  \multicolumn{5}{l}{For the latent disease severity} \\
  $\beta_0=-1$ & 0.007 & 0.114 & 0.945 & 0.114 \\
  $\beta_1=-0.2$ & $-$0.010 & 0.118 & 0.970 & 0.118 \\
  $\beta_2=0.8$ & 0.003 & 0.022 & 0.970 & 0.022 \\
  $\beta_3=-0.2$ & $-$0.001 & 0.015 & 0.940 & 0.015 \\
  $\sigma_1=1.5$ & 0.009 & 0.060 & 0.950 & 0.060 \\
  $\sigma_2=0.15$ & 0.000 & 0.007 & 0.960 & 0.007 \\
  $\rho=0.4$ & $-$0.003 & 0.048 & 0.935 & 0.048 \\
  \\
 \multicolumn{5}{l}{For the survival process} \\
  $\gamma=-0.12$ & $-$0.001 & 0.007 & 0.950 & 0.007 \\
  $\nu=0.75$ & 0.005 & 0.044 & 0.930 & 0.044 \\
  \\
  \multicolumn{5}{l}{For the first outcome (continuous)} \\
  $a_1=15$ & $-$0.035 & 0.471 & 0.955 & 0.471 \\
  $b_1=7$ & $-$0.024 & 0.183 & 0.960 & 0.184 \\
  $\sigma_\varepsilon=5$ & $-$0.000 & 0.099 & 0.960 & 0.099 \\
%  \\
  \multicolumn{5}{l}{For the second outcome (ordinal)} \\
  $a_{22}=1$ & 0.004 & 0.066 & 0.925 & 0.066 \\
  $a_{23}=2$ & 0.014 & 0.089 & 0.930 & 0.090 \\
  $a_{24}=4$ & 0.028 & 0.124 & 0.940 & 0.127 \\
  $a_{25}=5$ & 0.040 & 0.148 & 0.920 & 0.153 \\
  $a_{26}=6$ & 0.038 & 0.169 & 0.915 & 0.173 \\
%  \\
  \multicolumn{5}{l}{For the third outcome (ordinal)} \\
  $a_{31}=-1$ & 0.004 & 0.106 & 0.950 & 0.106 \\
  $a_{32}=1$ & 0.001 & 0.110 & 0.940 & 0.110 \\
  $a_{33}=3$ & 0.011 & 0.131 & 0.950 & 0.132 \\
  $a_{34}=4$ & 0.012 & 0.144 & 0.960 & 0.144 \\
  $a_{35}=6$ & 0.023 & 0.194 & 0.930 & 0.195 \\
  $a_{36}=8$ & 0.022 & 0.232 & 0.950 & 0.233 \\
  $b_3=1.2$ & $-$0.000 & 0.040 & 0.965 & 0.040 \\
   \hline
\end{tabular}
\end{table}

\begin{table}[h!tbp]
\caption{Area under the ROC curve (AUC) for the simulation study.} \label{tab:S4}
\centering
\begin{tabular}{rrcccccccc}
  \hline
   $t$ & $t'$ &  & Model 1 &  & Model JM &  & Cox &  & True AUC \\
  \hline
  3 & 9 &  & 0.922 &  & 0.909 &  & 0.892 &  & 0.934 \\
   & 12 &  & 0.920 &  & 0.908 &  & 0.875 &  & 0.943 \\
   & 15 &  & 0.915 &  & 0.903 &  & 0.853 &  & 0.952 \\
   & 18 &  & 0.907 &  & 0.896 &  & 0.830 &  & 0.959 \\
  \\
  6 & 9 &  & 0.926 &  & 0.911 &  & 0.883 &  & 0.930 \\
   & 12 &  & 0.930 &  & 0.915 &  & 0.868 &  & 0.940 \\
   & 15 &  & 0.932 &  & 0.916 &  & 0.847 &  & 0.950 \\
   & 18 &  & 0.930 &  & 0.914 &  & 0.825 &  & 0.958 \\
   \hline
\end{tabular}
\end{table}

\newpage
\section*{\cp{Stan} code for the simulation study}
{\scriptsize
\begin{verbatim}
data {
  int<lower=0> N_train; // Number of subjects in training data
  int<lower=0> obs; // Number of observations
  int subject[obs]; // Subject ID
  int<lower=0> K_ordi; // number of ordinal outcomes
  real Y_conti[obs];
  int<lower=0> Y_ordi[obs, K_ordi];
  int<lower=0> n_ordi; // Number of categories for ordinal outcomes
  vector[2] zero;
  real<lower=0> time[obs];
  int<lower=0> treat[obs];
  int<lower=0> treat_pts[N_train];
  int<lower=0, upper=100> age_pts[N_train];
  real tee[N_train]; // Survival time
  int<lower=0> event[N_train]; // Censoring indicator
}
parameters {
  vector<lower=-10, upper=10>[2] beta0;
  vector<lower=-10, upper=10>[2] beta1;
  vector[2] U[N_train];
  real<lower=0> var1;
  real<lower=0> var2;
  real<lower=-1, upper=1> rho;
  real<lower=0> var_conti;
  real gamma;
  real nu;
  real h0;
  real a_conti;
  real<lower=0> b_conti;
  real a_ordi_temp;
  real<lower=0> b_ordi_temp;
  vector<lower=0>[n_ordi-2] delta[K_ordi];
}
transformed parameters {
  real<lower=0> sig1;
  real<lower=0> sig2;
  cov_matrix[2] Sigma_U;
  real<lower=0> sd_conti;
  vector[n_ordi-1] a_ordi[K_ordi];
  vector<lower=0>[K_ordi] b_ordi;
  real theta[obs];
  real mu_conti[obs];
  real<lower=0, upper=1> psi[obs, K_ordi, n_ordi];
  vector<lower=0, upper=1>[n_ordi] prob_y[obs, K_ordi];

  // construct the latent variable theta
  for (i in 1:obs)
    theta[i] <- beta0[1] + beta0[2]*treat[i] + U[subject[i], 1] +
      (beta1[1] + beta1[2]*treat[i] + U[subject[i], 2])*time[i];

  // construct the means for the continuous variables
  for (i in 1:obs)
    mu_conti[i] <- a_conti + b_conti*theta[i];

  // construct the probability vector for the remaining ordinal variables
  a_ordi[1, 1] <- 0;
  for (l in 2:(n_ordi-1)) a_ordi[1, l] <- a_ordi[1, l-1] + delta[1, l-1] ;
  for (k in 2:K_ordi) {
    a_ordi[k, 1] <- a_ordi_temp;
    for (l in 2:(n_ordi-1)) a_ordi[k, l] <- a_ordi[k, l-1] + delta[k, l-1];
  }
  b_ordi[1] <- 1;
  for (k in 2:K_ordi) b_ordi[k] <- b_ordi_temp;

  for (i in 1:obs) {
    for (k in 1:K_ordi) {
      for (l in 1:(n_ordi-1)) {
        psi[i, k, l] <- inv_logit(a_ordi[k, l] - b_ordi[k]*theta[i]);
      }
      psi[i, k, n_ordi] <- 1;

      prob_y[i, k, 1] <- psi[i, k, 1];
      for (l in 2:n_ordi) {prob_y[i, k, l] <- psi[i, k, l] - psi[i, k, l-1];}
    }
  }

  sd_conti <- sqrt(var_conti);
  sig1 <- sqrt(var1);
  sig2 <- sqrt(var2);

  // construct the variance-covariance matrix
  Sigma_U[1,1] <- sig1*sig1;
  Sigma_U[1,2] <- rho*sig1*sig2;
  Sigma_U[2,1] <- Sigma_U[1,2];
  Sigma_U[2,2] <- sig2*sig2;
}
model {
  real h[N_train];
  real S[N_train];
  real LL[N_train];

  Y_conti ~ normal(mu_conti, sd_conti);
  for (i in 1:obs) {
    for (k in 1:K_ordi) {
      Y_ordi[i, k] ~ categorical(prob_y[i, k]);
    }
  }

  // construct random effects
  U ~ multi_normal(zero, Sigma_U);

  // construct survival part
  for (i in 1:N_train) {
    h[i] <- exp(gamma*age_pts[i] + nu*(beta0[1] + beta0[2]*treat_pts[i] + U[i, 1] +
                                         (beta1[1] + beta1[2]*treat_pts[i] + U[i, 2])*tee[i]))*h0;
    S[i] <- exp(-h0*exp(gamma*age_pts[i]+nu*(beta0[1]+beta0[2]*treat_pts[i]+U[i, 1])) *
                  (exp(nu*(beta1[1]+beta1[2]*treat_pts[i]+U[i, 2])*tee[i])-1) / (nu*(beta1[1]+beta1[2]*treat_pts[i]+U[i, 2])));
    LL[i] <- log(pow(h[i],event[i])*S[i]);  // event=1 for event; 0 for censored
  }
  increment_log_prob(LL);

  // construct the priors
  beta0 ~ normal(0, 10);
  beta1 ~ normal(0, 10);
  var1 ~ inv_gamma(0.01, 0.01);
  var2 ~ inv_gamma(0.01, 0.01);
  rho ~ uniform(-1, 1);
  var_conti ~ inv_gamma(0.01, 0.01);

  h0 ~ gamma(0.01, 0.01);
  nu ~ normal(0, 10);
  gamma ~ normal(0, 10);

  for (i in 1:(n_ordi-2)) delta[1, i] ~ normal(0, 10) T[0,] ;
  for (k in 2:K_ordi) {
    b_ordi_temp ~ uniform(0, 10);
    a_ordi_temp ~ normal(0, 10);
    for (i in 1:(n_ordi-2)) delta[k, i] ~ normal(0, 10) T[0,] ;
  }
}
\end{verbatim}
}

\section*{Full Conditionals}
For illustration purpose, we assume that there are one continuous outcome (denoted by $y_{i1}(t)$) and two ordinal outcomes (denoted by $y_{i2}(t)$ and $y_{i3}(t)$, respectively), while model~(3) is formulated as $\theta_{i}(t)=\boldsymbol{X}_{i}(t)\boldsymbol{\beta} + \boldsymbol{Z}_{i}(t)\boldsymbol{u}_i $. Assuming non-informative prior distribution for the parameter vector $\boldsymbol{\Theta}$, denoted by $f(\boldsymbol{\Theta})$, the joint likelihood is
\begin{eqnarray*}
  & & L(\boldsymbol{\Theta};\cdot) = p(\boldsymbol{y}|\boldsymbol{u}) p(\boldsymbol{u}) f(\boldsymbol{\Theta})  \\
  & \propto & \prod_{i=1}^I \bigg\{\prod_{j=1}^{J_i} p\big[Y_{i1}(t_{ij}) = y_{i1}(t_{ij})\big] p\big[Y_{i2}(t_{ij}) = y_{i2}(t_{ij})\big] p\big[Y_{i3}(t_{ij}) = y_{i3}(t_{ij})\big] \bigg\} \big\{h_i(t_i)^{\delta_i} S_i(t_i) \big\} p(\boldsymbol{u}_i) \\
  & = & \prod_{i=1}^I L_{y_1} L_{y_2} L_{y_3} L_S \cdot p(\boldsymbol{u}_i), \\
\end{eqnarray*}
where
\begin{flalign*}
L_{y_1} &= \prod_{j=1}^{J_i} \frac{1}{\sqrt{2\pi\sigma_\varepsilon^2}}\exp\bigg\{-\frac{\big[y_{i1}(t_{ij}) - a_1 - b_1\theta_i(t_{ij}) \big]^2}{2\sigma_\varepsilon^2} \bigg\}, && \\
L_{y_k} &= \prod_{j=1}^{J_i} \prod_{l=1}^{n_k} p\big[Y_{ik}(t_{ij})=l\big]^{I[Y_{ik}(t_{ij})=l]} && \\
&= \prod_{j=1}^{J_i} \prod_{l=1}^{n_k} \bigg\{p\big[Y_{ik}(t_{ij}) \le l|\theta_i(t_{ij})\big] - p\big[Y_{ik}(t_{ij}) \le l-1|\theta_i(t_{ij})\big] \bigg\}^{I[Y_{ik}(t_{ij})=l]}\\
&= \prod_{j=1}^{J_i}\bigg[\bigg\{1 - \textnormal{expit}\big[a_{k(n_k-1)} - b_k\theta_i(t_{ij})\big] \bigg\}^{I[Y_{ik}(t_{ij})=n_k]}  &&\\
 & ~~~~~ ~\cdot\prod_{l=2}^{n_k-1}\bigg\{\textnormal{expit}\big[a_{kl} - b_k\theta_i(t_{ij})\big] - \textnormal{expit}\big[a_{k(l-1)} - b_k\theta_i(t_{ij})\big]\bigg\}^{I[Y_{ik}(t_{ij})=l]} &&\\
 & ~~~~~ ~\cdot \bigg\{\textnormal{expit}\big[a_{k1} - b_k\theta_i(t_{ij})\big] \bigg\}^{I[Y_{ik}(t_{ij})=1]}\bigg],~k=2,3, &&  \\
L_S &= \bigg\{h_0(t_i)\exp\big[\boldsymbol{W}_i\boldsymbol{\gamma}+\nu \theta_i(t_i)\big]\bigg\}^{\delta_i} \exp\left[-\int_{0}^{t_i}h_0(s)\exp\big[\boldsymbol{W}_i\boldsymbol{\gamma}+\nu \theta_i(s)\big]ds\right], && \\
p(\boldsymbol{u}_i) &= \frac{1}{2\pi\sqrt{\boldsymbol{|\Sigma|}}}\exp\left[-\frac{1}{2}\boldsymbol{u}_i'\boldsymbol{\Sigma}^{-1}\boldsymbol{u}_i  \right], && \\
\textnormal{expit}(\cdot) &=
\frac{\exp(\cdot)}{1+\exp(\cdot)}. &&
\end{flalign*}

The full conditionals of all parameters are
\begin{enumerate}

\item  $f(a_1|\textnormal{others}) \propto N\left(\frac{\sum_{i=1}^{I}\sum_{j=1}^{J_i}\big[y_{i1}(t_{ij}) - b_1\theta_i(t_{ij})\big] }{N_T}, \frac{\sigma_\varepsilon^2}{N_T} \right);$

\item $f(b_1|\textnormal{others}) \propto N\left( \frac{\sum_{i=1}^{I}\sum_{j=1}^{J_i}\big[ y_{i1}(t_{ij}) - a_1\big ]\theta_i(t_{ij}) }{\sum_{i=1}^{I}\sum_{j=1}^{J_i} \theta_i(t_{ij})^2}, \frac{\sigma_\varepsilon^2}{\sum_{i=1}^{I}\sum_{j=1}^{J_i} \theta_i(t_{ij})^2} \right);$

\item $f(\frac{1}{\sigma_\varepsilon^2}|\textnormal{others}) \propto \textnormal{Gamma}\left( \frac{N_T}{2}+1, \frac{\sum_{i=1}^{I}\sum_{j=1}^{J_i}\big[y_{i1}(t_{ij}) - a_1 - b_1\theta_i(t_{ij})\big]^2 }{2} \right);$

\item $[\boldsymbol{a}_{2}, b_2 |\textnormal{others}] \propto \prod_{i=1}^I L_{y_2};$

\item $[\boldsymbol{a}_{3}, b_3 |\textnormal{others}]\propto \prod_{i=1}^I L_{y_3};$

\item $[\boldsymbol{\beta} |\textnormal{others}] \propto \prod_{i=1}^I L_{y_1} L_{y_2} L_{y_3} L_S;$

\item $[\boldsymbol{\gamma}, \nu |\textnormal{others}] \propto \prod_{i=1}^I L_S;$

\item $[\boldsymbol{u}_i|\textnormal{others}] \propto \bigg\{\prod_{j=1}^{J_i} p\big[Y_{i1}(t_{ij}) = y_{i1}(t_{ij})\big] p\big[Y_{i2}(t_{ij}) = y_{i2}(t_{ij})\big] p\big[Y_{i3}(t_{ij}) = y_{i3}(t_{ij})\big] \bigg\} L_S \cdot p(\boldsymbol{u}_i);$

\item $[\boldsymbol{\Sigma}|\textnormal{others}] \propto \prod_{i=1}^I p(\boldsymbol{u}_i),$

\end {enumerate}
where $N_T=\sum_{i=1}^{I}J_i$.

\end{document}